\documentclass[
  journal=pasa,
  manuscript=research-paper, 
  year=2023,
  volume=,
]{cup-journal}

\usepackage{microtype,siunitx,booktabs,physics,multicol,subfigure}
\usepackage{hyperref}
\hypersetup{colorlinks,citecolor=blue,linkcolor=blue,urlcolor=blue}
\makeatletter
\DeclareRobustCommand{\HI}{%
  \mbox{H\check@mathfonts\fontsize\sf@size\z@\selectfont I}%
}
\DeclareRobustCommand{\HII}{%
  \mbox{H\check@mathfonts\fontsize\sf@size\z@\selectfont II}%
}
\makeatother

\title{The MAGPI Survey: Drivers of kinematic asymmetries in the ionised gas of $z\sim0.3$ star-forming galaxies}

\author{R. S. Bagge} 
\affiliation{School of Physics, University of New South Wales, Kensington, NSW, 2032, Australia}
\alsoaffiliation{ARC Centre of Excellence for All Sky Astrophysics in 3 Dimensions (ASTRO 3D)}
\author{C. Foster} 
\affiliation{School of Physics, University of New South Wales, Kensington, NSW, 2032, Australia}
\alsoaffiliation{ARC Centre of Excellence for All Sky Astrophysics in 3 Dimensions (ASTRO 3D)}
\author{A. Battisti}
\affiliation{Research School of Astronomy and Astrophysics, Australian National University, Cotter Road, Weston Creek, ACT, 2611, Australia}
\alsoaffiliation{ARC Centre of Excellence for All Sky Astrophysics in 3 Dimensions (ASTRO 3D)}
\author{S. Bellstedt}
\affiliation{ICRAR,The University of Western Australia, 7 Fairway, Crawley, WA, 6009, Australia}
\author{M. Mun}
\affiliation{Research School of Astronomy and Astrophysics, Australian National University, Cotter Road, Weston Creek, ACT, 2611, Australia}
\alsoaffiliation{ARC Centre of Excellence for All Sky Astrophysics in 3 Dimensions (ASTRO 3D)}
\author{K. Harborne}
\affiliation{ICRAR,The University of Western Australia, 7 Fairway, Crawley, WA, 6009, Australia}
\alsoaffiliation{ARC Centre of Excellence for All Sky Astrophysics in 3 Dimensions (ASTRO 3D)}
\author{S. Barsanti}
\affiliation{Research School of Astronomy and Astrophysics, Australian National University, Cotter Road, Weston Creek, ACT, 2611, Australia}
\alsoaffiliation{ARC Centre of Excellence for All Sky Astrophysics in 3 Dimensions (ASTRO 3D)}
\alsoaffiliation{Sydney Institue for Astronomy, School of Physics, University of Sydney, NSW 2006, Australia}
\author{T.Mendel}
\affiliation{Research School of Astronomy and Astrophysics, Australian National University, Cotter Road, Weston Creek, ACT, 2611, Australia}
\alsoaffiliation{ARC Centre of Excellence for All Sky Astrophysics in 3 Dimensions (ASTRO 3D)}
\author{S. Brough}
\affiliation{School of Physics, University of New South Wales, Kensington, NSW, 2032, Australia}
\alsoaffiliation{ARC Centre of Excellence for All Sky Astrophysics in 3 Dimensions (ASTRO 3D)}
\author{S.M. Croom}
\affiliation{Sydney Institue for Astronomy, School of Physics, University of Sydney, NSW 2006, Australia}
\alsoaffiliation{ARC Centre of Excellence for All Sky Astrophysics in 3 Dimensions (ASTRO 3D)}
\author{C.D.P. Lagos}
\affiliation{ICRAR,The University of Western Australia, 7 Fairway, Crawley, WA, 6009, Australia}
\alsoaffiliation{ARC Centre of Excellence for All Sky Astrophysics in 3 Dimensions (ASTRO 3D)}
\author{T. Mukherjee}
\affiliation{Australian Astronomical Optics, Macquarie University, 105 Delhi Road, North Ryde, NSW 2113, Australia}
\author{Y. Peng}
\affiliation{Department of Astronomy, School of Physics, Peking University, Beijing 100871, China}
\alsoaffiliation{Kavli Institute for Astronomy and Astrophysics, Peking University, Beijing 100871, China}
\author{R-S. Remus}
\affiliation{Universit\"ats-Sternwarte, Fakult\"at f\"ur Physik, Ludwig-Maximilians-Universit\"at M\"unchen, Scheinerstr, 1, 81679, M\"unchen, Germany}
\author{G. Santucci}
\affiliation{ICRAR,The University of Western Australia, 7 Fairway, Crawley, WA, 6009, Australia}
\alsoaffiliation{ARC Centre of Excellence for All Sky Astrophysics in 3 Dimensions (ASTRO 3D)}
\author{P. Sharda}
\affiliation{Leiden Observatory, Leiden University, Postbus 9513, 2300 RA Leiden, The Netherlands}
\author{S. Thater}
\affiliation{Department of Astrophysics, University of Vienna, Türkenschanzstraße 17, 1180 Vienna, Austria}
\author{J. van de Sande}
\affiliation{Sydney Institue for Astronomy, School of Physics, University of Sydney, NSW 2006, Australia}
\alsoaffiliation{ARC Centre of Excellence for All Sky Astrophysics in 3 Dimensions (ASTRO 3D)}
\author{L. M. Valenzuela}
\affiliation{Universit\"ats-Sternwarte, Fakult\"at f\"ur Physik, Ludwig-Maximilians-Universit\"at M\"unchen, Scheinerstr, 1, 81679, M\"unchen, Germany}
\author{E. Wisnioski}
\affiliation{Research School of Astronomy and Astrophysics, Australian National University, Cotter Road, Weston Creek, ACT, 2611, Australia}
\alsoaffiliation{ARC Centre of Excellence for All Sky Astrophysics in 3 Dimensions (ASTRO 3D)}
\author{T. Zafar}
\affiliation{Australian Astronomical Optics, Macquarie University, 105 Delhi Road, North Ryde, NSW 2113, Australia}
\alsoaffiliation{Astronomy, Astrophysics and Astrophotonics Research Centre, Macquarie University, Sydney, NSW 2109, Australia}
\author{B. Ziegler}
\affiliation{Department of Astrophysics, University of Vienna, Türkenschanzstraße 17, 1180 Vienna, Austria}

\email[R. S. Bagge]{r.bagge@unsw.edu.au}

\doi{}

\received{dd Mmm YYYY}
\revised{dd Mmm YYYY}
\accepted{dd Mmm YYYY}
\published{2023}


\begin{document}

\begin{abstract}
Galaxy gas kinematics are sensitive to the physical processes that contribute to a galaxy's evolution. It is expected that external processes will cause more significant kinematic disturbances in the outer regions, while internal processes will cause more disturbances for the inner regions. Using a subsample of 47 galaxies ($0.27<z<0.36$) from the Middle Ages Galaxy Properties with Integral Field Spectroscopy (MAGPI) survey, we conduct a study into the source of kinematic disturbances by measuring the asymmetry present in the ionised gas line-of-sight velocity maps at the $0.5R_e$ (inner regions) and $1.5R_e$ (outer regions) elliptical annuli. By comparing the inner and outer kinematic asymmetries, we aim to better understand what physical processes are driving the asymmetries in galaxies. We find the local environment plays a role in kinematic disturbance, in agreement with other integral field spectroscopy studies of the local universe, with most asymmetric systems being in close proximity to a more massive neighbour. We do not find evidence suggesting that hosting an Active Galactic Nucleus (AGN) contributes to asymmetry within the inner regions, with some caveats due to emission line modelling. In contrast to previous studies, we do not find evidence that processes leading to asymmetry also enhance star formation in MAGPI galaxies. Finally, we find a weak anti-correlation between stellar mass and asymmetry (ie. high stellar mass galaxies are less asymmetric). We conclude by discussing possible sources driving the asymmetry in the ionised gas, such as disturbances being present in the colder gas phase (either molecular or atomic) prior to the gas being ionised, and non-axisymmetric features (e.g., a bar) being present in the galactic disk. Our results highlight the complex interplay between ionised gas kinematic disturbances and physical processes involved in galaxy evolution.
\end{abstract}

\section{INTRODUCTION }
\label{sec:int}
As galaxies evolve over time, signatures of their evolution remain in both the spatial distribution and kinematics of their ionised gas \citep{2016MNRAS.461..859S,2017MNRAS.471L..87O,2018MNRAS.476.4327L,2019MNRAS.483..458B,2022arXiv221009673J}. By studying these disturbances (or lack thereof) in the line-of-sight velocity distribution (LOSVD) of the emitting ionised gas, we can understand the physical mechanisms that contribute to galaxy evolution. Galaxies, however, rarely evolve in isolation and a longstanding goal of galaxy evolution studies is to understand the balance between internal and external galaxy formation processes that contribute to their evolution \citep[i.e., nature vs. nurture;][]{2009ApJ...690..102H}

The observed star-formation rates (SFRs) and cold gas masses in massive galaxies at $z=1-3$ suggest that galaxies should exhaust their supply of gas within a few Gyr \citep{2013ApJ...768...74T}, but a number of galaxies within the local universe (z $\sim$ 0) are still readily forming stars \citep{2008AJ....136.2846B,2010AJ....140.1194B}; this suggests that galaxies have a means of replenishing their cold gas supply to continue forming stars. This can be accomplished through several internal and external mechanisms. An example of an internal mechanism would be a star-forming galaxy's gravitational potential being deep enough to hold on to, and re-accrete the gas after it is expelled from either supernovae (SNe) or stellar winds; causing a galactic `fountain' effect \citep{1976ApJ...205..762S, 2008A&A...484..743S,2010MNRAS.404.1464M}. Conversely, rather than `recycling' expelled gas, galaxies can externally accrete fresh gas from the hot outer halo ($T \sim 10^6$ K), along cold ($T \sim 10^4$ K) cosmological filaments or through interactions and mergers with other galaxies \citep{2005MNRAS.363....2K,2009Natur.457..451D,2014A&A...567A..68D, 2017MNRAS.467..311P,2021MNRAS.504.5702W, 2022A&A...668A.182K}. For both the `cold' or `hot' modes, theory predicts that external gas accretion is a source of stellar-gas kinematic misalignments \citep[ie. difference between the rotation position angle (PA$_{kin}$) of the stars and gas;][]{2019MNRAS.483..458B}, although the internal properties of the galaxy affect whether the gas can realign or not \citep{2015MNRAS.448.1271L,2022MNRAS.514.2031C}. 

Gas is rarely stationary within the galactic disk, rotating with bulk motion as well as moving with a series of inflows and outflows. These gaseous flows onto the central regions of galaxies can trigger a galaxy's Active Galactic Nucleus (AGN), which can contribute to `quenching' the SFR in high stellar mass galaxies \citep[e.g.,][]{2007MNRAS.382.1415S}. AGNs are also a known source of outflows, and can disturb the kinematics of the entire galactic disk \citep[e.g.,][]{2011ApJ...732....9G,2016MNRAS.459.3144Z} and are a suspected cause of complex kinematic features in a velocity maps \citep{2022ApJ...925..203J}. 

The surrounding environment can have complex effects on a galaxy's SFR; ram-pressure stripping (RPS) caused by the hot intra-cluster medium and galaxy-galaxy interactions can readily quench the SFR in satellite galaxies through cold gas stripping \citep[see][for a review]{2021PASA...38...35C}, but can enhance the SFR in the most massive galaxy of a major merging pair \citep{2015MNRAS.452..616D}. In either case, these interactions can cause the kinematics to deviate significantly from disk-like rotation \citep{2018MNRAS.479..141T,2022MNRAS.515.3406M}.

Galaxy kinematics are extremely sensitive to both external and internal galaxy formation processes. By quantifying kinematic disturbances in galaxy velocity maps, we can better understand which physical processes dominate in their evolution. Integral Field Spectroscopic (IFS) instruments can collect spatially resolved spectra and can construct 2-dimensional kinematic maps of the stars and ionised gas phase in galaxies. In the last decade, IFS surveys (e.g., SINS; \citealt{2006ApJ...645.1062F}, ATLAS$^{\rm 3D}$; \citealt{2011MNRAS.413..813C}, SAMI; \citealt{2015MNRAS.447.2857B}, MaNGA; \citealt{2015ApJ...798....7B}, CALIFA;  \citealt{2016A&A...594A..36S}) have revealed the wide range of complex kinematic features and the disturbed kinematics that galaxies can display. These disturbances have been linked to external processes such as gas stripping and mergers \citep[e.g.,][]{2017ApJ...844...49B, 2018MNRAS.476.2339B, 2020ApJ...892L..20F, 2022MNRAS.517.2677R}, and internal processes such as star-formation, torques provided from a bar or the presence of an AGN \citep[e.g.,][]{2014A&A...568A..70B, 2015MNRAS.451.4397H, 2017MNRAS.465..123B,2020A&A...635A..41F}. 

The dynamics of stars and gas in galaxies are largely driven by the gravitational potential, and hence the mass distribution of the dark matter and baryons \citep[i.e., the stars and gas, see][]{2008gady.book.....B,2017ApJ...844...59B,2021ApJ...918...84R}, however \cite{2021MNRAS.505.3078V} show that galactic environment is also a significant driver. It is expected that external mechanisms such as galaxy-galaxy interactions or environmental effects (i.e., ram-pressure and viscous stripping) will produce a more significant kinematic disturbance in the outskirts where the gravitational potential is shallower than near the centre. This can be tested by comparing kinematic disturbances at inner and outer radial extents. A study using local, gas-rich galaxies in the \HI-SAMI survey \citep{2023MNRAS.519.1098C} compared the asymmetry measured in global atomic Hydrogen (\HI) and H$\alpha$ velocity profiles \citep{2023MNRAS.519.1452W}. The authors report that the global \HI\ asymmetry is driven mostly by external mechanisms (mergers and interactions), but did not find a straight-forward connection between the asymmetries of the two gas phases. 

In this work, we further explore the sources of kinematic asymmetry  by measuring the asymmetry confined to the optical disk by using a single gas phase (the ionised gas), rather than two different gas phases as was done in \cite{2023MNRAS.519.1452W}. We measure the asymmetry at the $0.5Re$ (\textit{inner}) and at $1.5Re$ (\textit{outer}) elliptical annuli, and investigate whether kinematic asymmetries vary with galactocentric radius. We use data from the Middle Ages Galaxy Properties in Integral field spectroscopy Survey \citep[MAGPI,][]{2021PASA...38...31F}, a Very Large Telescope/Multi-Unit Spectroscopic Explorer (VLT-MUSE) Large Program, to investigate whether the kinematics of galaxies do become more perturbed at larger radii. We use \textsc{kinemetry} \citep{2006MNRAS.366..787K} to quantify kinematic asymmetries as is commonly used in IFS surveys \citep{2008ApJ...682..231S,2011MNRAS.414.2923K,2017MNRAS.465..123B,2020ApJ...892L..20F}. The structure of this paper is as follows: \S\ref{sec:obs} gives a brief overview of the MAGPI survey, our data and sample selection. In \S\ref{sec::kinemetry} we describe the three \textsc{kinemetry} models used in this work. In \S\ref{sec::results} we discuss our results, the strength of each model and possible implications for galaxy evolution before offering some concluding remarks in \S\ref{sec::conc}.

We adopt a \cite{2020A&A...641A...1P} cosmology for this work; explicitly a flat Universe with $H_0$ = 67.7 kms$^{-1}$ Mpc$^{-1}$, $\Omega_{\rm M}$ = 0.31 and $\Omega_\Lambda$ = 0.69.

\section{DATA}
\label{sec:obs}
In this section, we briefly describe the MAGPI survey and the data products we use, followed by a description of our selection process and the subsample.

\subsection{The MAGPI Survey}
\begin{figure*}[htp]
  \centering
  \includegraphics[scale=0.55]{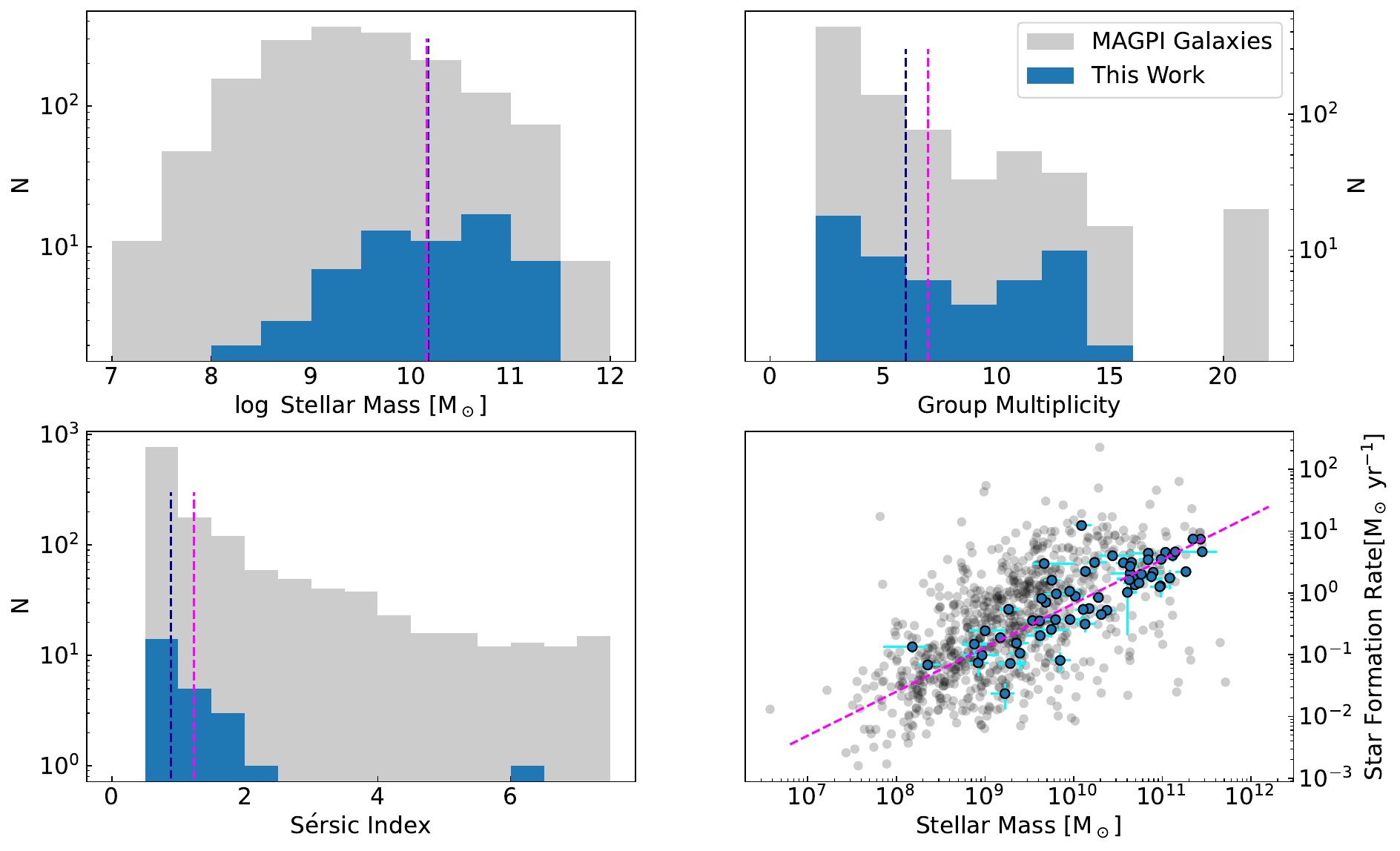}\quad
  \caption{Histograms of galactic properties for the parent MAGPI sample and the subsample used in this study. The parent MAGPI sample is shown in grey, whereas our subsample is shown in blue. Mean values for each property in the subsample are shown as magenta vertical lines in each panel. Median values for each property in the parent sample are shown as dark blue lines. \textit{Top Left}: Histogram of the stellar masses for the MAGPI galaxies and our sample. \textit{Top Right}: Histogram of the number of galaxies within the group containing the galaxy. \textit{Bottom Left}: Histogram of the S\'{e}rsic indices. Only galaxies with a $\chi^2$<10 from the surface-brightness fitting are shown in blue. \textit{Bottom Right}: Star formation rate vs. stellar mass. Cyan lines represents the errorbars in stellar mass and star formation rate. The magenta line shown is the fitted Main Sequence for the MAGPI galaxies at $z=0.35$.}
  \label{SFMS_sample}
\end{figure*}

\begin{figure*}[htp]
  \centering
  \includegraphics[scale=0.85]{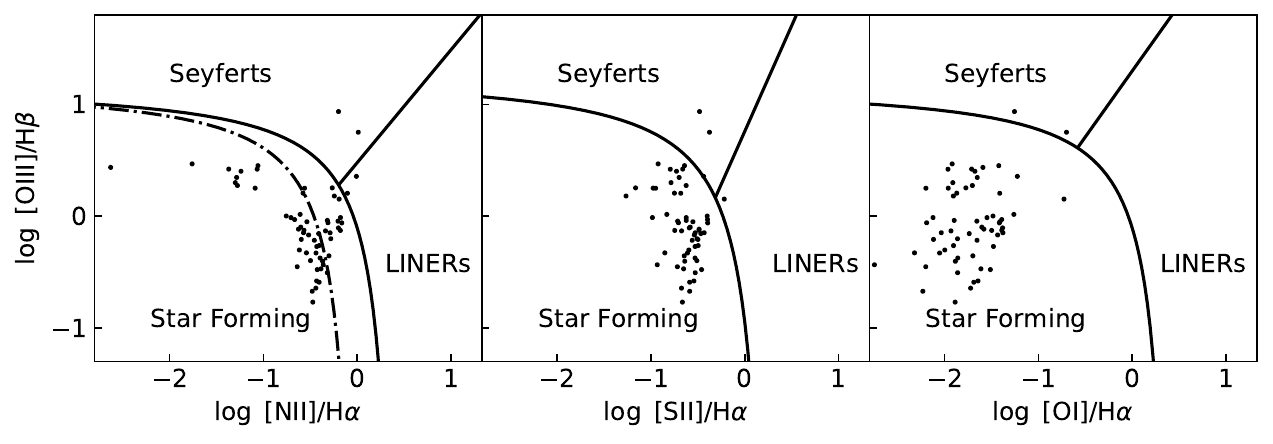}
  \caption{A [NII]-BPT (left), [SII]-BPT (middle) and [OI]-BPT (right) diagram of our sample. The hyperbolic solid lines in all three panels are from criterion suggested in \cite{2003MNRAS.346.1055K}. The dashed line in the left panel is the cutoff for \HII\ regions, while the solid line is the cutoff for AGNs. The straight line in the [NII]-BPT is the separation of Seyfert and LINERs suggested in \cite{2010MNRAS.403.1036C}. The straight lines in [SII]-BPT and [OI]-BPT separate Seyfert, LINER and \HII\ as per \cite{2006MNRAS.372..961K}.}
  \label{BPT}
\end{figure*}

The MAGPI survey is an ongoing Large Program on the VLT/MUSE targeting 56 $\sim$1$\times$1 arcmin fields (currently 41 completed). MAGPI data are spatially resolved spectra of stars and ionised gas focusing on the redshift range $z= 0.28 - 0.35$ selected from three of the Galaxy and Mass Assembly \citep[GAMA,][]{2011MNRAS.413..971D} fields: G12, G15 and G23. MAGPI also includes archival MUSE observations from the legacy fields Abell 370 and Abell 2477. When observations are completed, the main MAGPI sample will comprise spatially resolved spectroscopy for 60 `primary' targets (M$_*$ $\geq$ 7$\times$10$^{10}$ M$_\odot$) and $\sim$ 100 spatially resolved satellites (M$_*$ $\geq$10$^9$ M$_\odot$). A detailed description of the survey and its aims are discussed in \citep{2021PASA...38...31F}.

Data are taken with the MUSE Wide Field Mode in the nominal wavelength range (4650 \AA\ - 9300 \AA) with a spectral sampling of 1.25 \AA. Ground-layer adaptive optics is used to mitigate the effects of atmospheric seeing, resulting in a 270 \AA-wide gap between 5780 \AA\ and 6050 \AA\ due to the GALACSI laser notch filter. Each MAGPI field has a field-of-view of 1x1 square arcminutes with spatial sampling of 0.2$''$ pixel$^{-1}$ and the average image qaulity of 0.65$''$ FWHM. A detailed discussion of the data reduction will be presented in the first data release (Mendel et al. in prep.).

There are a number of high-level data products currently available to the MAGPI team. Stellar masses for this paper are derived via broadband photometry SED fitting, using 9-band u-Ks imaging data collated for the GAMA survey. To derive the photometric measurements for MAGPI sources, these images were first pixel-matched to the MAGPI cubes (ensuring that the pixel scales were identical), and then the MAGPI segmentation maps presented by \citet{2021PASA...38...31F} were used to extract forced photometry per object. These photometry were then used for SED fitting with the code \textsc{ProSpect} \citep{2020MNRAS.495..905R} to provide a range of galaxy properties. These stellar masses are extracted in the standard approach as done in the most recent GAMA analysis  \citep{2020MNRAS.498.5581B,2022MNRAS.513..439D}, and in \citet{2023MNRAS.522.3602D}. A truncated skewed-Normal parametrisation is used for the star-formation history, with a linearly evolving metallicity evolution implemented to ensure a chemical build-up that follows the stellar mass build-up in the galaxy. A \cite{2003PASP..115..763C} Initial Mass Function (IMF) is assumed, and the \cite{2003MNRAS.344.1000B} stellar population templates are used.

We also use S\'{e}rsic indices computed from 2-dimensional surface brightness fitting using \textsc{galfit} \citep{2002AJ....124..266P,2010AJ....139.2097P}. We model the two-dimensional surface brightness distribution in mock \emph{i}-band images computed from the MAGPI cubes. We adopt a model of the reconstructed point spread function (PSF) derived using \texttt{muse-psfr} \citep{2020A&A...635A.208F} as described by Mendel et al., (in prep.). In addition to the primary galaxy being fitted (i.e., the galaxy of interest), we include in our fit neighbouring galaxies within $\Delta m_i = 5$ mag and projected separations $R_p < 5(R_\mathrm{e,primary} + R_\mathrm{e,neighbour})$. All other sources in the image are masked according to the segmentation image generated by \textsc{ProFound} \citep{2018MNRAS.476.3137R}. Before fitting we estimate and remove the local sky background around each source using neighbouring sky pixels (i.e., pixels outside of the dilated \textsc{ProFound} segment mask in the \emph{i}-band image). Our choice of using MAGPI mock \emph{i}-band images for deriving S\'{e}rsic indexes is motivated by the MAGPI data themselves providing the most uniform coverage in terms of depth, image quality and availability of ancillary data.

We also use emission line products, in particular continuum subtracted fluxes and kinematic maps, that are described in Battisti et al. (in prep.), Briefly, the emission line products were derived using GIST (Galaxy IFU Spectroscopy Tool; \citealt{2019A&A...628A.117B}), a \textsc{python} wrapper of \textsc{pPXF} \citep{2004PASP..116..138C,2017MNRAS.466..798C} and \textsc{GandALF} \citep{2006MNRAS.366.1151S,2006MNRAS.369..529F}. Gas velocity and dispersion maps are derived from single component Gaussian fits to spectral lines in the continuum subtracted spectra ranging from [OII]3727\AA\ to [SII]6733\AA; velocity and dispersion measurements for all lines are then tied to brightest line (H$\alpha$ for our redshift range). We use H$\alpha$ fluxes measured from the Gaussian fit to calculate a SFR within 1.5$R_e$ for each of the galaxies in our sample. We correct for dust attenuation assuming an intrinsic Balmer decrement of H$\alpha$/H$\beta$ = 2.86 and the extinction curve from \cite{1989ApJ...345..245C} with a reddening factor $R_V$=3.1. We use the H$\alpha$ SFR calibration from \cite{2013seg..book..419C}, which implicitly assumes a \cite{2001MNRAS.322..231K} IMF. Following \cite{2010MNRAS.404.2087B}, we multiply our SFR calibration by 1.22 to correct for the \cite{2001MNRAS.322..231K} IMF and continue with a \cite{2003PASP..115..763C} IMF for our subsequent analysis.

For environment metrics for MAGPI galaxies, we use results from the group-finding algorithm, `Parliment'\footnote{A parliment is the collective noun for a group of magpies} (Harborne et al., in prep). In particular, the comoving distance to the nearest neighbour $d_1$ in the groups that Parliment found in each of the MAGPI fields. Parliment follows the methodology in \cite{2009ApJ...697.1842K}. Briefly, galaxies were determined to be within a group if the angular and line-of-sight separation were less than the perpendicular and parallel linking lengths, respectively. The parallel and perpendicular linking lengths for galaxy $i$ were calculated as:
\begin{equation}
    l_{\perp,i} = min\left(L_{max}\times(1+z_i),\frac{b}{\bar{n}^{(1/3)}}\right)
    \label{perpendicular_linking_length}
\end{equation}
and
\begin{equation}
    l_{\parallel,i} = R\ l_{\perp,i}
    \label{parallel_linking_length}
\end{equation}
where $\bar{n}=0.01$ Mpc$^{-3}$ was the assumed number density of galaxies \citep{2019MNRAS.490.1451F} and $L_{max}$, $b$ and $R$ are the free parameters of the algorithm.

\label{subsec:MAGPI}

\subsection{Sample Selection}
We select those galaxies from the GAMA fields G12, G15 and G23 such that they are within MAGPI's primary redshift range (0.28$<z_{\rm spec}<$0.35) and have an effective radius $R_e$ (ie. radius containing half the $i$-band light from the galaxy) larger than 0.7$''$, which is 0.05$''$ larger than the estimated PSF in each MAGPI field. Our redshift and size cutoff ensures we only target galaxies at this predicted `epoch of transformation' and are sufficiently resolved. Finally, we remove those galaxies where the maximum H$\alpha$ flux SNR$<$20. This reduces the entire MAGPI catalogue to 61 galaxies. If a galaxy satisfies these criteria, we mask any spaxels on the velocity map were the H$\alpha$ flux SNR$<$3.

\label{subsec:sample}
For our parent sample of 61 galaxies, most galaxies in our sample sit along the Star Forming Main Sequence \citep[SFMS;][]{2007ApJ...670..156D,2014ApJS..214...15S,2021MNRAS.505..540T} with 39 galaxies within $\pm$0.3 dex of the SFMS (see bottom right panel of Fig. \ref{SFMS_sample}) defined for MAGPI galaxies at $z\sim0.35$ (Mun et al., in prep). It is not surprising that our sample mostly consists of star-forming galaxies, since one selection criterion we invoked was a sufficient detection of H$\alpha$ in emission. Looking at S\'{e}rsic indices only for galaxies where the fit has converged (i.e., $\chi^2$<10), these 29 galaxies (48\% of the parent sample) consists mostly of `disky' galaxies with the mean and median 2D S\'{e}rsic indices $n$ being 1.27 and 0.9, respectively (e.g., bottom left panel Fig. \ref{SFMS_sample}). The sample consists of mostly massive galaxies ($\log$ (M$_*$/M$_\odot$)>10) with a mean and median of 10.17 and 10.18, respectively (e.g., top left panel Fig. \ref{SFMS_sample}). Looking at the environments the galaxies in our sample inhabit, they inhabit low to high-density environments. The median and mean group multiplicity of our sample is 6 and 7, respectively (e.g., top right panel of Fig. \ref{SFMS_sample}).

Finally, we used BPT diagnostic plots \citep{1981PASP...93....5B} and the criteria stipulated in \cite{2003MNRAS.346.1055K} and \cite{2006MNRAS.372..961K}, shown in Fig. \ref{BPT}, to classify which galaxies in our sample are star-forming, host an AGN, or have \HII+AGN ionisation. To be consistent with our calculated SFRs, we use integrated fluxes within 1.5$R_e$, masking spaxels where the SNR$<$3 for each line. For a galaxy to be classified as a Seyfert or a Low Ionisation Nuclear Region (LINER), we require consistent classification across the three [NII]-BPT, [SII]-BPT and [OI]-BPT plots, and it not be in the \HII+AGN section of the [NII]-BPT. Our sample consists of star-forming galaxies (43/61; 70\%), galaxies with \HII+AGN ionisation (16/61; 26\%)\ and a small number of Seyferts (2/54; 3\%). No LINERs were detected consistently across all three BPT diagrams, but one is detected using the [SII]-BPT diagram.

\begin{table*}
\begin{tabular}{lccccccccl}
\toprule
MAGPIID & $z$ & Stellar Mass & $L_{H\alpha}$ & Ionisation Source & Central/Satellite & $N_{Group}$ & Asymmetric? & $v_{\rm asym} (1.5R_e) > v_{\rm asym}(0.5R_e)$  \\
 & & [x10$^{10}$ M$_\odot$]  & [x10$^{40}$ erg s$^{-1}$]& & & & & \\
\midrule
(1)&(2)&(3)&(4)&(5)&(6)&(7)&(8)&(9) \\
\bottomrule
1201302222 & 0.299 & 0.149 & 8.308 & \HII& Cen & 2 & Asymmetric & TRUE \\ 
1202197197 & 0.291 & 10.849 & 170.086 & \HII & Cen & 2 & Not Asymmetric & TRUE \\ 
1203076068 & 0.306 & 2.744 & 97.565 & \HII & Sat & 10 & Not Asymmetric & TRUE \\ 
... & ... & ... & ... & ... & ... & ... & ... & ... \\ 
\bottomrule
\end{tabular}
\caption{\textbf{Col. 1}: MAGPIID, \textbf{Col. 2}: spectroscopic redshift, \textbf{Col. 3}: Stellar Mass, \textbf{Col. 4}: Integrated H$\alpha$ luminosity within $2R_e$, \textbf{Col. 5}: Source of ionisation in the galaxy, \textbf{Col 6.}: Is the galaxy the Central or a satellite to the group? A dsah indicates no data was available, \textbf{Col 7.}: Number of galaxies within the group containing the galaxy, a dash indicates that no data was available, \textbf{Col 8.}: Is the galaxy asymmetric  globally? (ie. $\langle v_{\rm aysm} \rangle>0.04$), \textbf{Col 9.}: Does the galaxy have higher asymmetric outskirts? (ie. $v_{\rm asym} (1.5R_e) > v_{\rm asym}(0.5R_e)$). Dashes indicate where data was not available. The remaining rows of the table are provided as supplementary material.}
\label{Table1}
\end{table*}

\section{Kinemetric Analysis}
\label{sec::kinemetry}
\begin{figure*}
  \centering
  \subfigure{\includegraphics[scale=0.55]{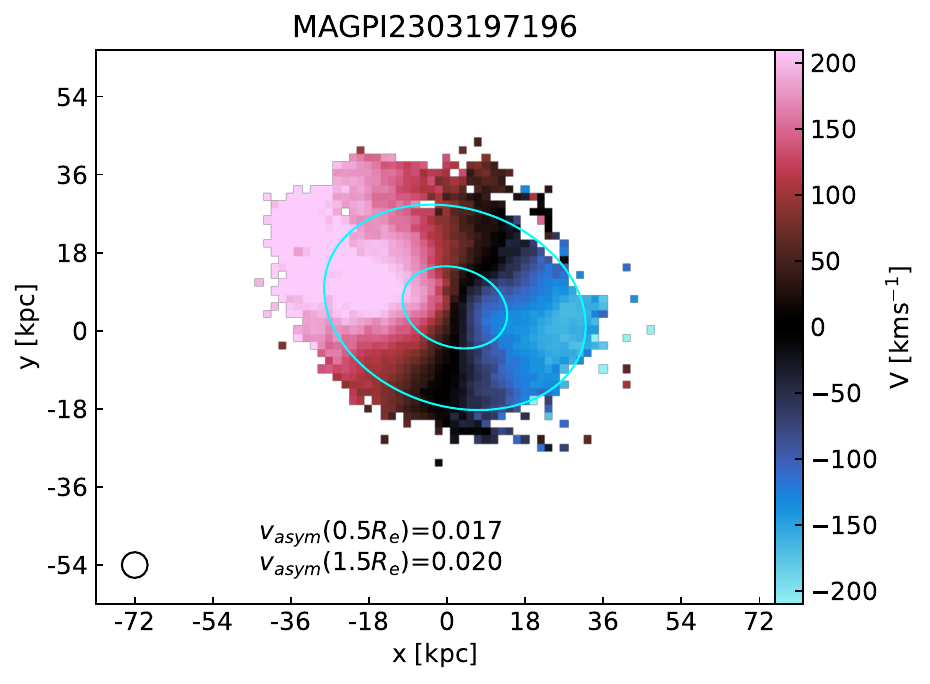}}
  \subfigure{\includegraphics[scale=0.55]
  {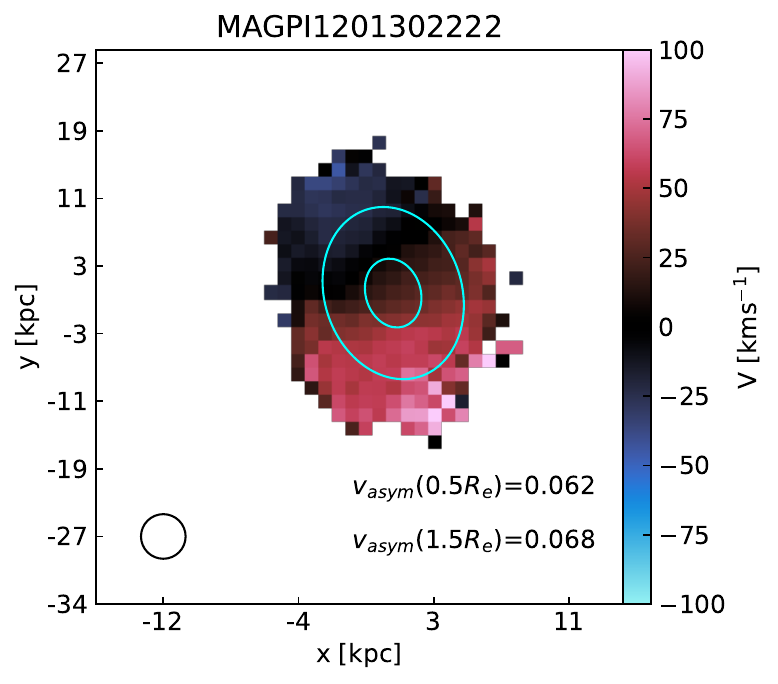}}
  \caption{Gas velocity maps of a symmetric (\textit{left}) and asymmetric (\textit{right}) galaxy, where an asymmetric galaxy is defined as $\langle v_{\rm asym} \rangle>0.04$. MAGPI2303197196 is a rather normal rotating galaxy, whereas MAGPI120130222 is a slower-rotating galaxy with a complicated velocity structure. The cyan ellipses correspond to $0.5R_e$ and $1.5R_e$, respectively. The estimated PSF for each galaxy is shown as a circle in the lower left corner.}
  \label{Complex_Normal}
\end{figure*}
\textsc{kinemetry} \citep{2006MNRAS.366..787K} is a generalisation of photometric techniques for higher order moments of the LOSVD \citep[e.g., mean velocity, disperision, $h_3$, $h_4$;][]{1993MNRAS.265..213G,1993ApJ...407..525V}. These kinematic moments can be symmetric, with either \textit{even} parity (point-symmetric) or \textit{odd} parity (point-antisymmetric); \textsc{kinemetry} fits a model to the LOSVD moment maps using a series of concentric ellipses with position angle (PA) and axial ratio $q$, where $q=b/a$ and $b$ and $a$ are the semi-minor and semi-major axes, respectively. Similar to other tilted ring fitting algorithms \citep{1974ApJ...193..309R,1994ApJ...436..642F,2007ApJ...664..204S}, \textsc{kinemetry} fits a Fourier Series along the ellipse as follows:
\begin{equation}
    K(a,\theta) = A_0 + \sum_{N=1}^{N=n} [A_n\sin{(n\theta)} + B_n\cos{(n\theta)}],
    \label{FS}
\end{equation}
where $a$ is the semi-major axis of the ellipse, $\theta$ is the azimuth along the ellipse with respect to the semi-major axis, A$_0$ is the zeroth harmonic term and A$_n$ and B$_n$ are the $n\rm{th}$ additional harmonic terms. \textsc{kinemetry} determines the best fitting ellipses by minimising the harmonic coefficients that are in the series. Eqn. \ref{FS} can be equivalently represented as: 
\begin{equation}
    K(a,\theta) = A_0 + \sum_{n=1}^{n=N} k_n\cos{(n[\theta - \phi_n(a)])},
    \label{compactFS}
\end{equation}
where $k_n = \sqrt{A_n^2+B_n^2}$ and $\phi = \arctan\frac{A_n}{B_n}$. One can extract the $k_n$ values, which describes the kinematics of the galaxy, and the PA and $q$ parameters, which describe the geometry of the ellipse. If a galaxy is rotating with perfect circular motion, it can be completely described using Eqn. \ref{FS} with a single cosine term, which would be symmetrical in the azimuthal axis at 180$^{\circ}$. Any deviations from circular motion would result in asymmetric function, and these asymmetries would be encoded as power (ie. a non-zero values in the A$_n$ and B$_n$ terms where $n>1$) in the higher-order coefficients in Eqn. \ref{FS} (see Fig. \ref{ellipse_plots} for an example of what these asymmetries look like). The first moment of the LOSVD (the recessional velocity map) is an odd moment, and it was suggested by \cite{2006MNRAS.366..787K} that only odd harmonic terms ($k_1,k_3,k_5$) in Eqn. \ref{compactFS} are needed to sufficiently describe ellipses along this moment. The subsequent moment following the recessional velocity map (the dispersion map) is an even moment of the LOSVD; and rather than only odd harmonic terms, both even \textit{and} odd harmonic terms are needed. For the purposes of this work, we only focus on the first moment of the LOSVD, i.e. the recessional velocity map. 

\textsc{kinemetry} has been previously used on a number of IFS surveys to quantify the asymmetry in the stellar velocity and dispersion maps. The ATLAS$^{3D}$ survey \citep{2011MNRAS.413..813C} used \textsc{kinemetry} to describe their sample as either `regular' or `non-regular' rotators (RR, NRR) by whether the luminosity-weighted average $\langle{k_5/k_1}\rangle$ reached above 0.04 \citep{2011MNRAS.414.2923K}. A similar criterion was adopted for the SAMI Galaxy Survey \citep{2015MNRAS.447.2857B,2017MNRAS.465..123B,2017MNRAS.472.1809B}, where those with $\langle{k_5/k_1}\rangle < 0.04$ were described as RR, those with $\langle{k_5/k_1}\rangle$ between 0.04 and 0.08 as quasiregular rotators and those with $\langle{k_5/k_1}\rangle > 0.08$ as NRRs \citep{2021MNRAS.505.3078V}. For this paper, we adopt a definition of a galaxy being `asymmetric' if $v_{\rm asym}>0.04$ \citep{2011MNRAS.414.2923K}. 

\textsc{kinemetry} is agnostic to the source of the emission and can be used on velocity maps of both the stars and gas in galaxies. The Spectroscopic Imaging survey in the Near-infrared with SINFONI \citep[SINS;][]{2006ApJ...645.1062F} survey defined the asymmetry in the H$\alpha$ velocity field as,
\begin{equation}
    v_{\rm asym,SINS} =\frac{k_2+k_3+k_4+k_5}{4k_1},
    \label{SINS}
\end{equation}
where $v_{\rm asym}$ is averaged over the entire disk.The units of $k_n$ will be dependent on the map being modelled; in our case for a velocity map, the units are in kms$^{-1}$, hence $v_{\rm asym}$ is a dimensionless quantity. They included the even terms as major mergers are expected to cause large kinematic disturbances, and so even terms would be needed to adequately model the data. They found that $v_{\rm asym,SINS}$ was a reliable proxy to distinguish disk and merging systems \citep{2008ApJ...682..231S}. In contrast, \cite{2017MNRAS.465..123B,2017MNRAS.472.1809B} choose to only include the odd terms and compute the mean of the ratio of the higher order to first order moments over all radii ($v_{\rm asym}$) as follows:
\begin{equation}
    v_{\rm asym,SAMI} = \frac{k_3+k_5}{2k_1}.
    \label{SAMI}
\end{equation}
Their choice of only including odd terms was due to negligible power being distributed to the even terms for their sample. They found that kinematically asymmetric galaxies scattered below the Tully-Fisher Relation \citep[TFR;][]{1977A&A....54..661T} and that kinematic asymmetry in the gas is inversely correlated with stellar mass \citep{2002AJ....123.2358K,2014ApJ...795L..37C, 2017MNRAS.472.1809B}.

For this study, we focus on measuring the kinematic asymmetry present in a sample of MAGPI galaxies at $0.5R_e$ and $1.5R_e$ to study how the kinematic disturbances change with galactocentric distance. We tested three different \textsc{kinemetry} models: 
\begin{itemize}
\item{\textbf{M1:}} the first model includes only odd terms (ie. $k_1,k_3,k_5$) and the centre fixed on the brightest pixel;
\item{\textbf{M2:}} the second model includes odd and even terms (ie. $k_1,k_2,k_3,k_4,k_5$) and the centre fixed on the brightest pixel; and finally,
\item{\textbf{M3:}} the third model includes only odd terms (ie. $k_1,k_3,k_5$), but the position of the centre is allowed to vary. 
\end{itemize}

For models M1 and M3, we define the asymmetry following Eqn. \ref{SAMI}, and for model M2, we define the asymmetry following Eqn. \ref{SINS}. We use the v$_{\rm asym}$ values fitted at $0.5R_e$ and $1.5R_e$ applying equal weighting to each spaxel along the ellipse. Across all three models, we initially set the kinematic position angle and axis ratio to coincide with their photometric counterparts (i.e. $PA_{\rm kin}=PA_{\rm phot}$). The initial kinematic centre is assumed to be located at the coordinates of the brightest pixel in the MUSE $i$-band. We set parameter boundaries as follows: $PA_{kin}$=[0,360] degrees and $q_{\rm kin}$=[$q_{\rm phot}$-0.1,$q_{\rm phot}$+0.1]. In addition, and for M3 only, we allow the position of the kinematic centre to vary freely as determined by \textsc{kinemetry}. We use \textsc{kinemetry} with a set of ellipses between $0.5R_e$ and $1.5R_e$ with the distance in-between each ellipse equal to half the estimated seeing ($\sim0.3''$). Uncertainties on $v_{\rm asym}$ are estimated using 100 Monte Carlo realisations, where we re-run \textsc{kinemetry} on the velocity map with Gaussian noise (corresponding to the uncertainty in the velocity measurement) injected to each pixel. The final $v_{\rm asym}$ values and uncertainties correspond to the mean and standard deviation of the Monte Carlo distribution. We also calculate a flux-weighted $\langle v_{\rm asym} \rangle$ averaged over 1.5$R_e$. While we mask spaxels on the velocity maps where H$\alpha$ SNR<3, this leads to discontinuities in the maps that prevent \textsc{kinemetry} from computing asymmetries, we therefore estimate velocity value of these spaxels by taking the median of the adjacent 8 spaxels. See \ref{app1} for details.

Before conducting any analysis, we remove three galaxies where we are unable to measure the asymmetry at $1.5R_e$, we also remove ten galaxies where the uncertainty in the asymmetry at $1.5R_e$ is larger than 0.2 and where we have replaced more than 30\% of the spaxels along the $1.5R_e$ ellipse (See Fig. \ref{v_asym_SNR}). These choices in cutoff are necessarily ad-hoc, but we found that this limit prevents us from removing low asymmetry galaxies with small uncertainty and removes galaxies with uncertain values that are due to poor fits from \textsc{kinemetry}, GIST or the fraction of missing data is to high. Galaxies with small $R_e$ and low SNR are naturally excluded as there are too few spaxels along the ellipse for \textsc{kinemetry} to perform the fitting. Five of the ten galaxies thus removed are \HII+AGN galaxies with compact size and low H$\alpha$ SNR. A further Seyfert galaxy is also removed. We include the other AGN and \HII+AGN galaxies where the kinematics are modelled adequately according to the uncertainties in their asymmetry following that expected for purely star-forming galaxies.

One galaxy (MAGPI1207197197) had been excluded from any kinemetric analysis due to a poor fit from GIST, but it is mentioned briefly in Sect. \ref{environment} due to it being in the same group as two galaxies in our sample. Examples of typical symmetric and asymmetric galaxies are shown in Fig. \ref{Complex_Normal}. The final sample consists of 47 galaxies

\section{Results and Discussion}
\label{sec::results}
\begin{figure*}
    \centering
    \subfigure{\includegraphics[scale=0.52]{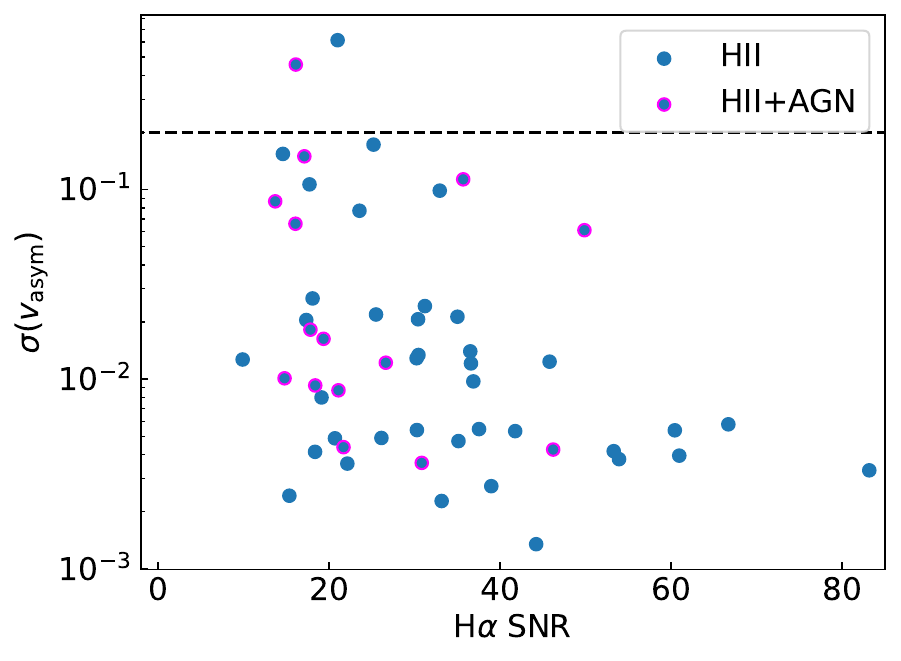}}
    \subfigure{\includegraphics[scale=0.52]{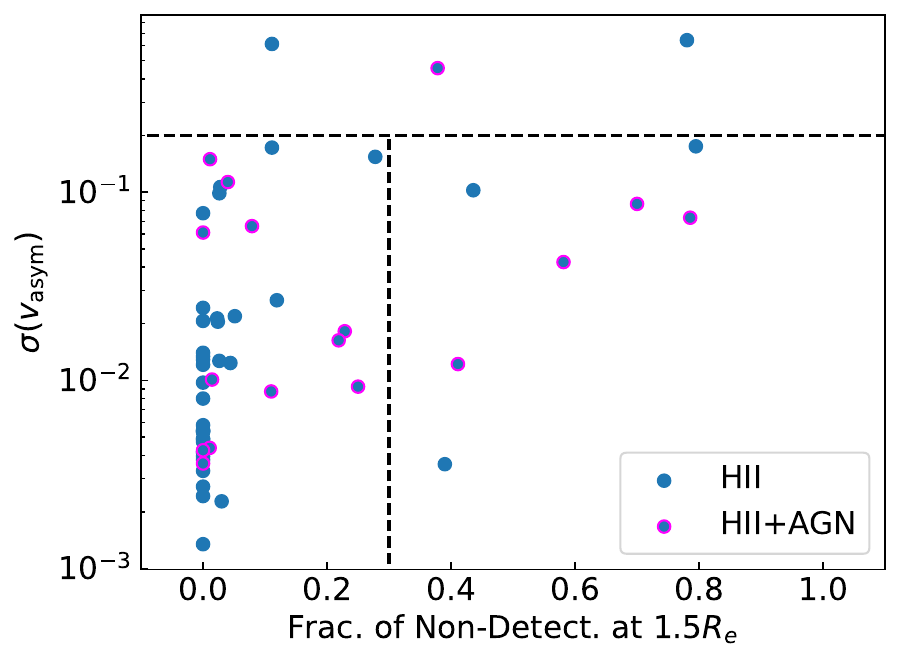}}
    \caption{\textit{Left}: Uncertainty in $v_{asym}$ against the average SNR at that ellipse. AGN galaxies have a magenta edges. When calculating the SNR, we run \textsc{kinemetry} on the SNR map and fix the ellipses to same PA and $q$ values that are returned when running \textsc{kinemetry} on the velocity map. The dashed black line is our adoptive cutoff in uncertainty to remove galaxies from our sample. \textit{Right}: Uncertainty in $v_{asym}$ against the fraction of non-detected spaxels at the 1.5$R_e$ ellipse. As well as excluding galaxies where the uncertainty is larger than 0.2, we also remove galaxies with more than 30\% missing data along the ellipse.}
    \label{v_asym_SNR}
\end{figure*}

In this section, we investigate the connection between asymmetry and galactic properties of each of the galaxies in our sample. In particular, their environment probed through projected distance to nearest neighbour, whether they host an AGN, and their star-formation activity and stellar mass. 

We conduct our analysis using asymmetries calculated with M2. We discuss, in detail, our motivation behind this choice in \ref{app1}. Using M2, 81$^{+6}_{-4}$\%\footnote{The uncertainties on all fractions, henceforth, are calculated following the methodology in \cite{2011PASA...28..128C}} of our sample displays larger asymmetry in their outskirts compared to their inner regions, and 46$^{+6}_{-7}$\% are asymmetrical at all radii, and a KS-test suggests that the distributions of $v_{asym}(1.5R_e)$ and $v_{asym}(0.5R_e)$ values are different distributions ($t=0.34,p<0.009$). The right panel in Fig. \ref{v_asym_SNR} shows a histogram of $v_{asym}(0.5R_e)$ and $v_{asym}(1.5R_e)$. Both of these results support the hypothesis that a galaxy's outer kinematics are more likely to be disturbed. We now discuss possible physical drivers of this.

\begin{figure}
    \centering
    \includegraphics[scale=0.5]{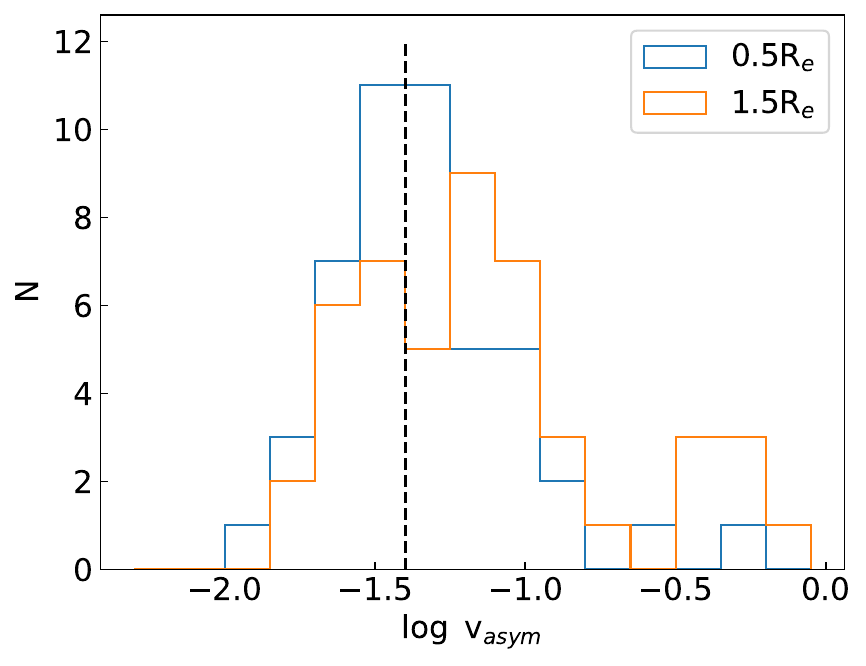}
    \caption{Histograms of $\log v_{asym}$ at $0.5R_e$ (blue) and $1.5R_e$ (orange). The black dashed represents our asymmetry cutoff ($\log v_{asym} =-1.39$). Both histograms peak at the same values ($\log v_{asym}\sim -1.35$), though the asymmetry values at $1.5R_e$ are skewed to higher values. A KS-test suggest that we can be confident at the 99.9\% level ($t=0.34,p<0.009$) that the each values are drawn from different distributions. The large fraction of galaxies with more asymmetric outskirts and the KS-test results supports our hypothesis that galaxy kinematics are more disturbed at larger galactocentric radius}
    \label{vasym_hist}
\end{figure}

\subsection{Environment}
\label{environment}
The asymmetry in velocity fields is usually attributed to external processes like mergers and interactions \citep{2008ApJ...682..231S,2018MNRAS.476.2339B,2020ApJ...892L..20F} or from ram-pressure stripping from the ICM \citep{2008A&A...483..783K,2010A&A...520A.109K}. For our current subsample, we only have environmental measurements for 42 galaxies. The environmental metrics for MAGPI galaxies are currently incomplete due to limited field of view from MUSE (1$''$x1$''$), compared to the large GAMA fields they are targeting. Of the 42 galaxies, 15 galaxies are centrals (ie. the brightest galaxy in the group) with the remaining 27 being satellites. Six central galaxies (5/15; 33\%$^{+9\%}_{-13\%}$) are considered globally asymmetric, whereas 20 satellites are globally asymmetric (16/27; 59\%$^{+9\%}_{-8\%}$).

As two galaxies begin to interact with each other, the outer regions would be more disturbed compared to the inner regions. Interestingly, the fraction of satellite galaxies that are more asymmetric in their outskirts is much larger than centrals with more asymmetric outskirts (93\%$^{+3\%}_{-2\%}$ vs. 66\%$^{+13\%}_{-6\%}$). This suggests that satellites are much more susceptible to disturbed outskirts than central galaxies.

Although interactions between centrals and satellites offers a neat explanation for those groups with both asymmetric centrals and satellites, it does not explain the why the fraction of asymmetric centrals is small while the fraction of asymmetric satellites is large. \cite{2022MNRAS.515.3406M} found that for interacting galaxies with a 2:5 stellar mass ratio in the Feedback In Realistic Environments Simulations \citep[FIRE;][]{2014MNRAS.445..581H,2018MNRAS.480..800H}, $v_{\rm asym}$ in the more massive galaxy will peak during the first pass of the interaction but will return to values that are similar to that of an isolated disk galaxy, before peaking again and remaining at high values after coalescence. It could be that the asymmetric satellites in our sample have not made, or have just made, their first passage, hence the central galaxy appears to be symmetric. Rather then being driven by a global group property such as group multiplicity or halo mass, our findings suggest that kinematic asymmetries are likely driven by smaller scale properties, such as projected distance to neighbouring galaxies. This was found to be the case in \cite{2020ApJ...892L..20F}, who found that the fraction of galaxies with high asymmetry increases when the projected distance between galaxies in close pairs decreases.

\subsubsection{Projected distance to nearest neighbour}
\begin{figure}
  \centering
  \includegraphics[scale=0.5]{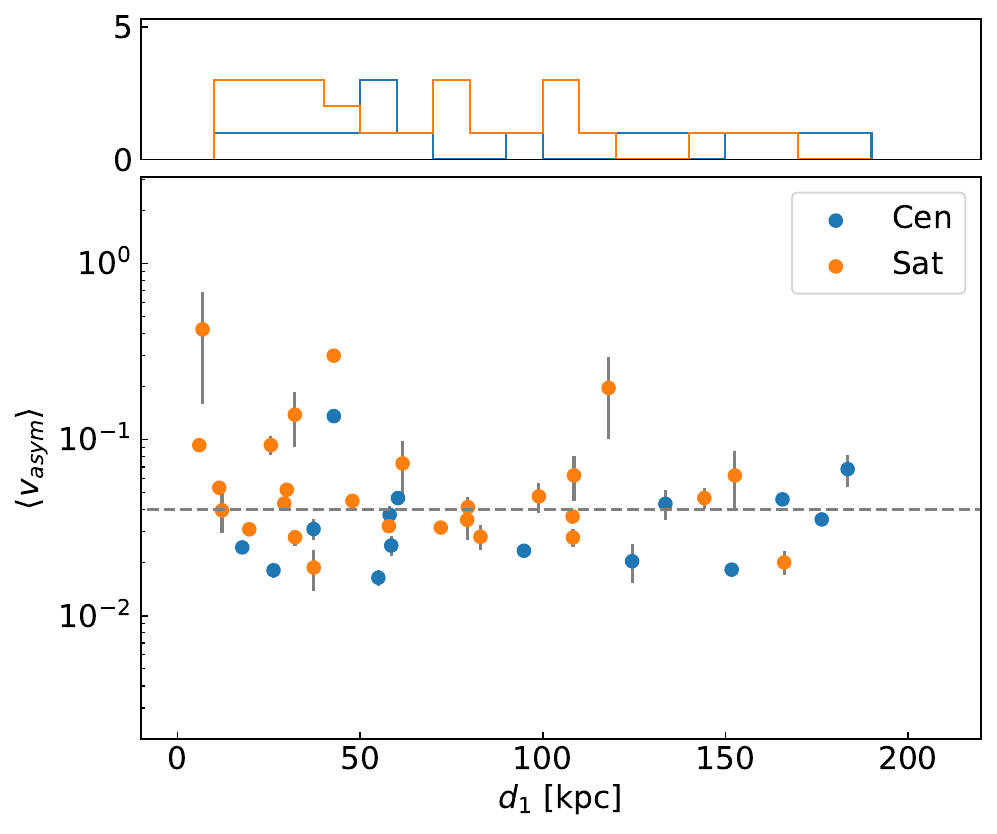}
  \caption{$\langle v_{asym} \rangle$ vs. distance to the nearest neighbour, $d_1$ where the points are coloured by whether they are a central (blue) or satellite (satellite) galaxy. The grey dashed line is the asymmetry cutoff. Galaxies are, on average, more asymmetric as the projected distance to the nearest neighbour decreases; however this is primarily found in satellite galaxies, not centrals. The histogram above shows the number of asymmetric central and satellite galaxies within the $d_1$ bins of 10 kpc.}
  \label{d1}
\end{figure}
Having a large projected distance between galaxies is also a likely explanation for those groups with satellites that are symmetric, as they are probably further away (but still within linking distance) and are not interacting with the central. This can be seen in Fig. \ref{d1} where we plot $\langle v_{asym} \rangle$ against the co-moving projected distance to the nearest neighbouring galaxy. We see that the asymmetry increases as the projected distance decreases. This effect is more severe in satellite galaxies, where the fraction of asymmetric satellites increases as the projected distance decreases, whereas the fraction of asymmetric centrals stays roughly the same (See the histogram above Fig. \ref{d1}). This is consistent with the findings in \cite{2022MNRAS.515.3406M} where the central galaxy became, and stayed asymmetric, after coalescence. It should also be said that \cite{2022MNRAS.515.3406M} investigate asymmetries in the \textit{stellar} LOSVD, and not the ionised gas, as we have done in this work. 

\cite{2018MNRAS.476.2339B} used ionised gas kinematic asymmetries, as we have done in this work and found that the distance to the nearest neighbour had a stronger influence on the asymmetry, than the distance to the 5$^{th}$ neighbouring galaxy. Our results, alongside \cite{2018MNRAS.476.2339B}, would suggest that the symmetric satellites and centrals with small projected distances have most likely not experienced their first pericenter passage, as was found in \cite{2022MNRAS.515.3406M}.

\begin{figure*}
    \centering
    \includegraphics{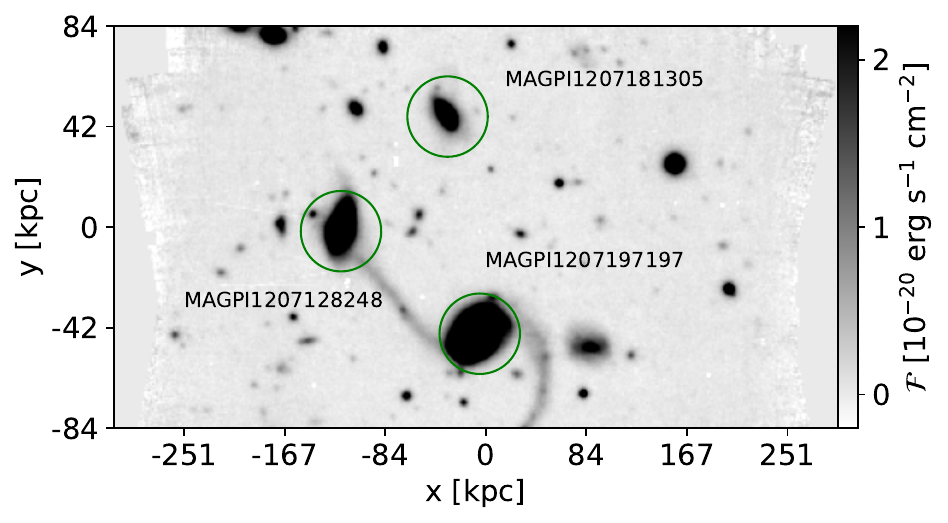}
    \caption{A \emph{i}-band cutout of the MAGPI1207 field, the three galaxies with green circles are MAGPI1207197197 (bottom right), MAGPI1207128248 (centre left), MAGPI1207181305 (above). Multiple other sources are in the field, but are unlabelled as they do not belong to the group featured. What is extremely noticeable is the presence of a tidal tail connecting MAGPI1207197197 and MAGPI1207128248 and the extremely extended, loosely wound spiral arm or a tidal tail.}
    \label{MAGPI1207}
\end{figure*}

There is an example of one group consisting of two asymmetric satellites around a central galaxy. The group containing MAGPI1207197197, MAGPI1207128248, MAGPI1207181305 is a clearly interacting system with visible streams and tails connecting the individual galaxies (Fig. \ref{MAGPI1207}). This is consistent with findings in \cite{2020ApJ...892L..20F} as the galaxies are clearly interacting with each other and are in close proximity (e.g., MAGPI1207197248 has a projected distance $\sim109$ kpc and MAGPI1207181305 has a projected distanc $\sim138$ kpc. Unfortunately, there were no other examples of all the galaxies in a single group being asymmetric, since not all of the galaxies within an individual group satisfied our selection criteria. 

\subsubsection{Asymmetry in previous gas phase}
We now hypothesise a scenario for the asymmetric outskirts we see in our sample. If we assume we are primarily tracing \HII\ regions (i.e., formation of OB stars), we might expect that the ionised gas will retain some memory of its kinematic state from before it is ionised, when it was either atomic or molecular. It should be noted that ionised gas can also be found in a diffuse component \citep[Diffuse Ionised Gas, DIG;][]{1990ApJ...349L..17R} since it is expected that DIG can account for as much as $\sim$ 50 \% of the H$\alpha$ emission in galaxies \citep{2007ApJ...661..801O, 2016ApJ...827..103K}. It is also likely that the surrounding DIG would also inherit the same kinematic state of the \HII\ region since DIG is primarily from ionizing photons `leaking' out of nearby \HII\ regions with a mean free path of $\sim $ 2 kpc \citep{2022A&A...659A..26B}. Gas can be ionised through physical processes other than star formation (e.g., AGN, shocks), we discuss these in the Sect. \ref{AGN}.

The now ionised gas would not move far from where it was when it was in the previous phase, due to the near-instantaneous snapshot the H$\alpha$ line provides; meaning that that the asymmetry we observed in the ionised gas may reflect that of the previous atomic or molecular gas phase. The now ionised gas would maintain the same kinematic disturbance from its previous phase. Having asymmetric outskirts in the ionised gas would be consistent if it was actually due to the \HI\ becoming disturbed first, either from ram-pressure stripping or interactions. Findings of disturbed morphologies and kinematics for \HI, CO (tracing molecular gas) and spatially coinciding H$\alpha$ `blobs' lends support for a galaxy's environment being responsible for the disturbed kinematics in the ionised gas \citep{2017MNRAS.466.1382L,2022ApJ...936..133C}. 

It should be said that this is only true if the \HI\ ends up closer to optical disk after the environmental interaction, due to \HI\ reaching a larger radial extent compared to the molecular and ionised gas. Though recent studies have found that the environment can impact \HI\ within the \HI\ truncation radius, $R_{\HI}$, as environmental stripping lowers the \HI\ gas surface density of the entire galaxy \citep{2023arXiv230307549W}. It seems reasonable that a decreasing gas surface density (ie. a redistribution of the gas) would be accompanied by a kinematic disturbance.

The relationship between the asymmetries in ionised and atomic phases was investigated more deeply in \cite{2023MNRAS.519.1452W} where they report that the connection was more nuanced then initially thought. However, their analysis of asymmetry was restricted to global flux profiles, not asymmetries in velocity maps. Analysing global flux profiles would wash out all the asymmetries, which we find to localised to specific regions. Using \textsc{kinemetry} on resolved \HI\ velocity maps may result in a stronger connection between the ionised gas and atomic phase.

\subsection{Active Galactic Nuclei}
\label{AGN}
Since $\sim$ 29\% of the galaxies in our sample host an AGN or \HII+AGN ionisation, we can investigate if an AGN can explain galaxies with larger inner asymmetries compared to their outskirts. AGNs are proven essential for reproducing the high mass end of the observed stellar mass function by preventing hot halo gas from cooling, effectively stopping further star formation in massive galaxies \citep{1998A&A...331L...1S,2006MNRAS.370..645B,2006ApJS..163....1H, 2015MNRAS.446..521S, 2017MNRAS.465...32B}. AGN are also often found in galaxies with disturbed kinematics  \citep{2011ApJ...732....9G,2016MNRAS.459.3144Z,2019MNRAS.487.2491E,2022ApJ...925..203J}. Being centrally located within the galaxy, we might expect that the asymmetry would be larger \textit{inside} a galaxy than it is in the outskirts for a galaxy with an AGN. The nuclear region of an AGN galaxy can have a misaligned PA compared to their outskirts \citep[e.g.,][]{2022MNRAS.516.1442I, 2022MNRAS.517.2677R}, this would cause larger asymmetries when performing \textsc{kinemetry} on nuclear regions.

There are caveats in our analysis with AGN and \HII+AGN galaxies that should be mentioned before we discuss our results. First, the kinematic maps used in this analysis are constructed from a set of Gaussian fits to spectral lines where the width and relative velocity are tied to one line; H$\alpha$ for our redshift range. When we observe galaxies with \HII+AGN ionised gas components, these are combined in the LOSVD. This can be an issue when fitting single component Gaussians to the spectra \citep{2013MNRAS.436.2576L,2015ApJ...806...84L,2017ApJ...834...30F}. This implies that the kinematic maps modelled may not be adequate in the nuclear region. Kinematic maps of the central regions that use multiple components might provide more informative kinematics of these regions, but fitting emission with multiple components is a complex problem that requires high spectral resolution, high signal-to-noise and adequate software, which is beyond the scope of this paper. It is also possible that the inner regions are unresolved for galaxies whose $0.5R_e$ is only just larger than our spatial resolution cutoff, consequently the complex kinematic signatures would be smoothed out from resulting beam smearing. To mitigate these effects, AGN galaxies with poor GIST fits have been removed from this analysis (see Fig. \ref{v_asym_SNR} and the second last paragraph of Sect. \ref{sec::kinemetry}), and we have only included those AGN galaxies where the uncertainties are no worse than expected for star-forming galaxies.

\begin{figure*}
  \centering
  \subfigure{\includegraphics[scale=0.50]{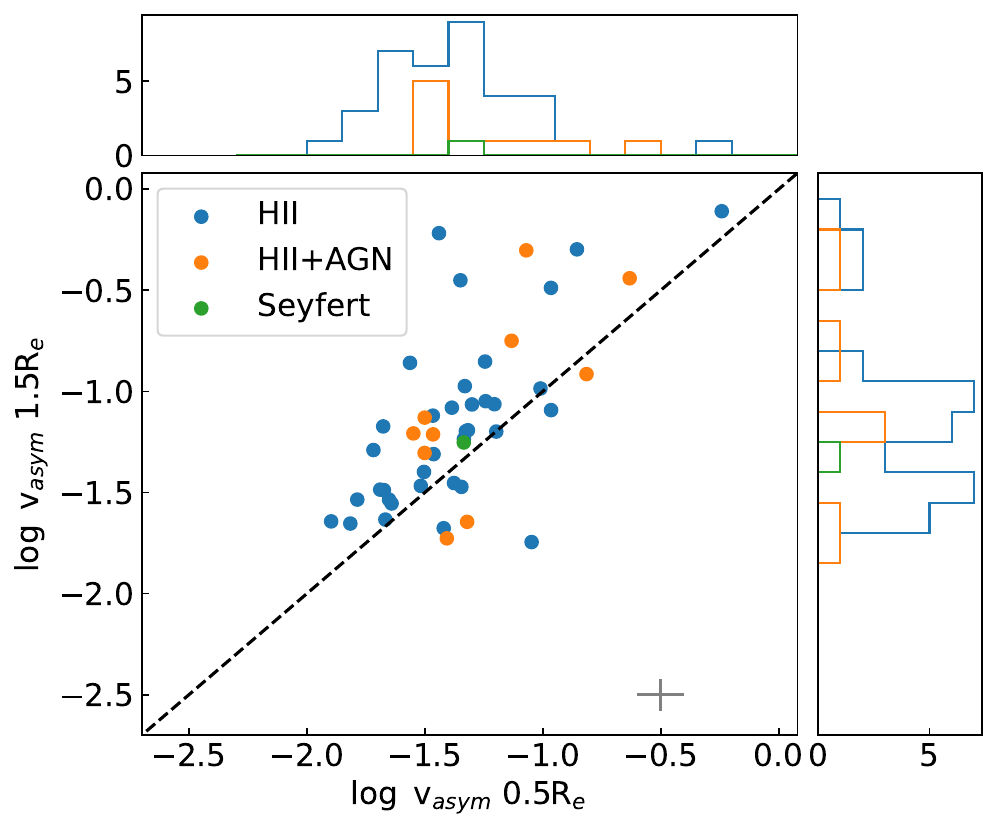}}
  \subfigure{\includegraphics[scale=0.50]{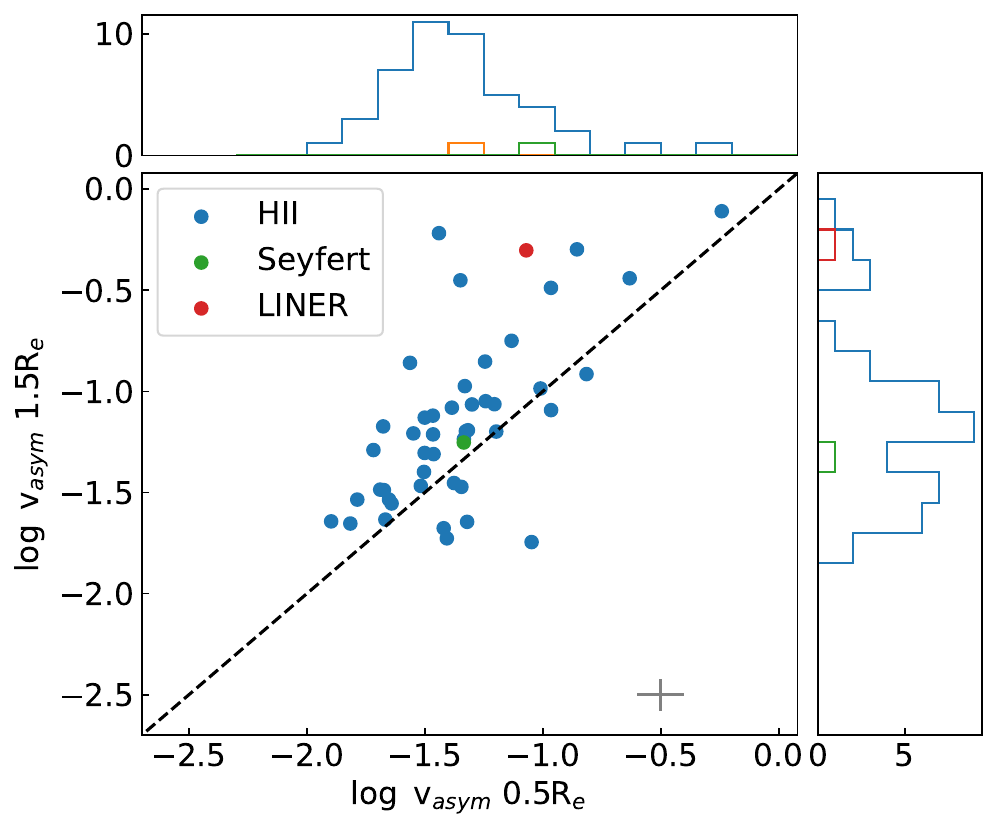}}
  \caption{\textit{Left}: v$_{asym}$ (1.5R$_e$) vs. v$_{asym}$ (0.5R$_e$) where the points are coloured according to the classification from Fig \ref{BPT}. \textit{Right}: The same are \textit{left}, except the points are coloured by the classification according to the [SII]-BPT in Fig. \ref{BPT}. The dashed line is both is 1:1. The median errorbars for the sample is shown in the bottom right corner. The histograms above and the side of each shows the distribution of v$_{asym}$ of each ionisation source. \textit{Right}: The same as \textit{left}, except we use only the classification from [SII]-BPT. It is included to break the degeneracy between purely AGN and \HII+AGN galaxies.}
  \label{AGNs}
\end{figure*}

The left panel of Fig. \ref{AGNs} shows $v_{\rm asym}$ ($0.5Re$) vs. $v_{\rm asym}$ ($1.5Re$) for each classification across all three BPT diagrams in Fig. \ref{BPT}. The Seyfert galaxy and seven of the ten \HII+AGN galaxies have $v_{\rm asym} (1.5R_e)>v_{\rm asym} (0.5R_e)$. Using the [SII]-BPT can break some of the degeneracy between purely AGN and \HII+AGN \citep[e.g.,]{2006MNRAS.372..961K}. To that end, we investigate the same $v_{\rm asym}$ ($0.5Re$) vs. $v_{\rm asym}$ ($1.5Re$) parameter space using the classification from the [SII]-BPT diagram (See right panel of Fig. \ref{AGNs}). Although there are only two confirmed AGNs using [SII]-BPT, both have larger asymmetries at 1.5$R_e$. Looking at the histograms above and to the right of the panels in Fig. \ref{AGNs}, we see that those galaxies with AGN or \HII+AGN have larger asymmetries on average than purely star-forming galaxies. This suggests that hosting an AGN is associated, on average, with larger kinematic asymmetries in both the inner and outer regions.

Our results contrast with those in \cite{2017MNRAS.465..123B}, who found no evidence to suggest that an AGN was the cause of $v_{\rm asym}$. They are, however, consistent with findings in\HI-SAMI where there was a lack of a connection between the asymmetry measured from \HI\ global flux distributions (which primarily traces asymmetries at larger radial extents) and the more centrally concentrated ionised gas which come from AGN \citep{2023MNRAS.519.1452W}. The impact of AGN have on velocity maps in simulated galaxies has shown that those with AGN demonstrated larger asymmetries, and that velocity maps of simulated galaxies where AGN feedback was not implemented were inconsistent with the asymmetric AGN galaxies in CALIFA \citep{2020A&A...635A..41F}.

It is hard to draw conclusions given the small number of confirmed AGN galaxies in this study (one Seyfert and one LINER using the [SII]-BPT). Given that most of the non-star-forming galaxies are \HII+AGN, it could be that AGN, or some other processes are creating shocks, rather than outflows, leading to kinematics disturbances in the ionised gas at 1.5$R_e$ \citep{2014MNRAS.442..784Z,2016MNRAS.461.3111B}. However, if these AGN galaxies have outflow signatures present in their spectra, the kinematic disturbances are most likely due to the outflows from the AGN.


\subsection{Star-Formation and Stellar Mass}
In this section, we discuss the possible impact the star formation activity and stellar mass could have in driving the asymmetry in the galaxies in our sample. 
\begin{figure*}
  \centering
  \subfigure{\includegraphics[scale=0.50]{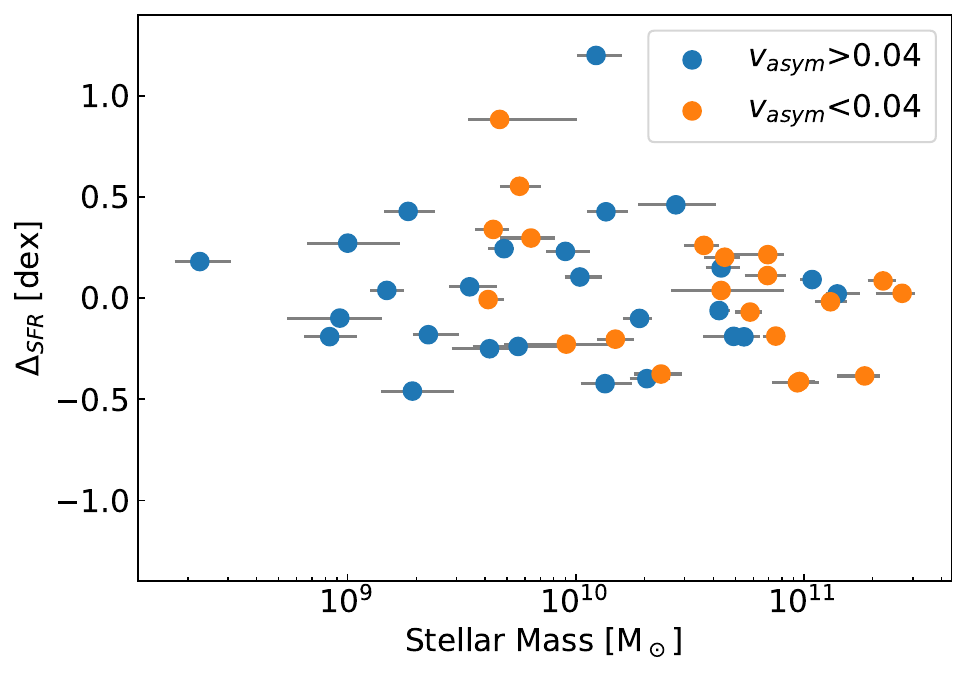}}
  \subfigure{\includegraphics[scale=0.50]{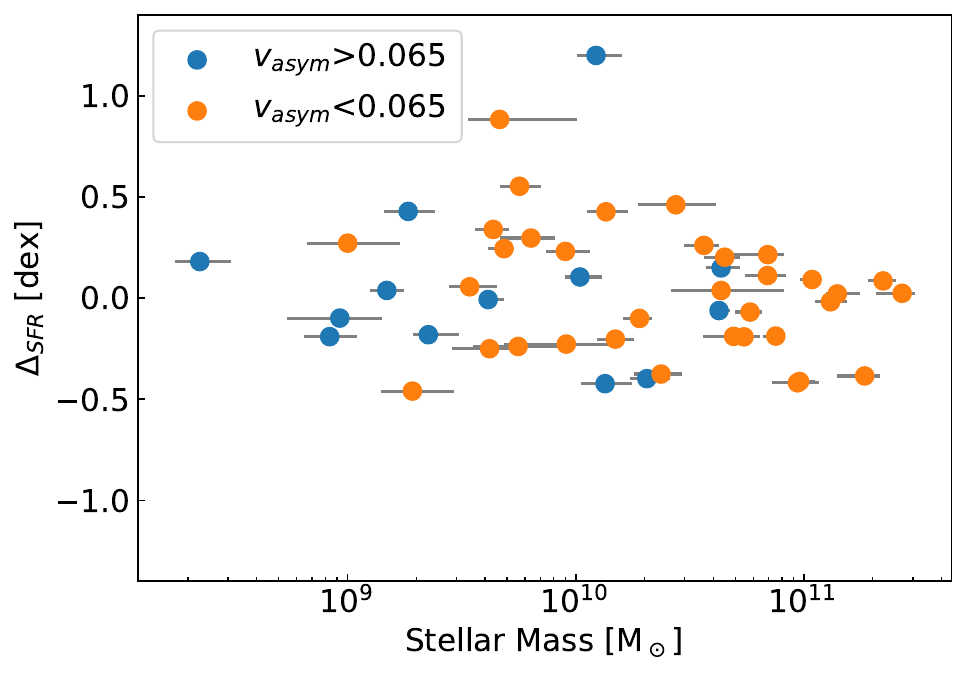}}
  \caption{\textit{Left}: Logarithmic distance from SFMS ($\Delta_{\rm SFR}$) vs. Stellar Mass for MAGPI galaxies. Asymmetric galaxies (e.g., $v_{asym}>0.04$) are coloured blue, whereas symmetric galaxies ($v_{asym}<0.04$) are coloured orange. \textit{Right}: The same as \textit{left} except we now use the same methodology for calculating $v_{\rm asym}$, and cuttoff as \cite{2017MNRAS.465..123B}. Including even terms in the asymmetry measure leads to more asymmetric, main sequence galaxies.}
  \label{SFMS_asym}
\end{figure*}

\subsubsection{Star Formation}
Since a large asymmetry value is a result of the galaxy displaying more non-circular motion relative to its rotational motion (characterised by $k_1$), we can investigate if more non-circular motion results in less rotational support in the gas. Less rotational support, from the non-circular motions, in the gas would lead to more gas collapsing, and we might expect the now ionised gas to inherit the same non-circular motions. To characterise whether our galaxies are experiencing enhanced star-formation due to the asymmetry, we consider the Star Forming Main Sequence (SFMS, Fig. \ref{SFMS_asym}) to determine which galaxies are showing increased or decreased star-formation activity. The SFMS for MAGPI galaxies at $z\sim0.35$ is given by $\log_{10}$(SFR) = 0.712$\pm0.04\times\log_{10}$(M$_*$) - 7.293$\pm$0.06 (see Mun et al., in prep. for more information). We adopt a conservative logarithmic difference between the estimated SFR and the SFMS of $\Delta_{SFR}>$0.5 for star-bursting galaxies, 0.5$>\Delta_{SFR}>$-1 for main sequence galaxies, and $\Delta_{MS}<$-1 for quenched galaxies. We find that there are three star bursting galaxies, and no quiescent galaxies.

We find that only one of the three star-bursting galaxies in our sample is asymmetric. We find that at low stellar masses, main sequence galaxies in our sample tend to be more asymmetric than their high stellar mass counterparts (See the left panel of Fig. \ref{SFMS_asym}). This result is consistent with those found in the MaNGA survey \citep{2020ApJ...892L..20F}; which found that at low stellar mass ($\log$ (M$_*$/M$_\odot$) < 10) star-bursting and main sequence galaxies displayed larger values of asymmetry compared to quiescent galaxies; but at large stellar masses, quiescent galaxies were either more, or at least as asymmetric as star-bursting and main sequence galaxies. Since we do not have any quiescent galaxies in our sample, we do not comment on the asymmetry in quiescent galaxies.

We also compare our results with those in \cite{2017MNRAS.465..123B}. To do this we adopt their methodology; chiefly, the larger cutoff value in asymmetry (0.065 vs. 0.04) and re-computing $\langle v_{\rm asym} \rangle$ using only odd terms, although we stress that these terms must be considered (See \ref{app1}). Using their cutoff and $v_{asym}$ value, our results are consistent with \cite{2017MNRAS.465..123B}. We again find that only one star-bursting galaxy is considered asymmetric, and the majority of our main sequence galaxies are symmetric (See right panel of Fig. \ref{SFMS_asym}). Comparing our results between the different methodologies would suggest that the inclusion of even terms increases the asymmetry in main sequence galaxies. 

There are some caveats for the comparisons between SAMI, MaNGA and MAGPI that must be mentioned; the sample size used in \cite{2017MNRAS.465..123B,2017MNRAS.472.1809B} and \cite{2020ApJ...892L..20F} is much larger than ours ($\sim$350 vs.$\sim$50) meaning our sample could be missing a number of asymmetric and symmetric galaxies at all stellar masses. MaNGA and SAMI also average over smaller radial extents to our sample (MaNGA integrates out to 1$R_e$ and SAMI only uses 3 ellipses for all galaxies, regardless of physical size). As we have shown in \ref{app1} $v_{\rm asym}$ will more often be larger at more extended radii, causing our asymmetry values to be larger. However, considering we use a luminosity weighted asymmetry which increases the weighting towards the brighter, inner regions where the asymmetry will be smaller, this should be a minor affect. Thus, even with these caveats we are confident that the differences between MaNGA, SAMI and MAGPI (ie. more asymmetric galaxies at all stellar masses in MAGPI) are real. Whether this is evolution of kinematic disturbances due to MAGPI being at higher redshift, will require a further analysis that is beyond scope of this paper.

\begin{figure}
    \centering
    \includegraphics[scale=0.55]{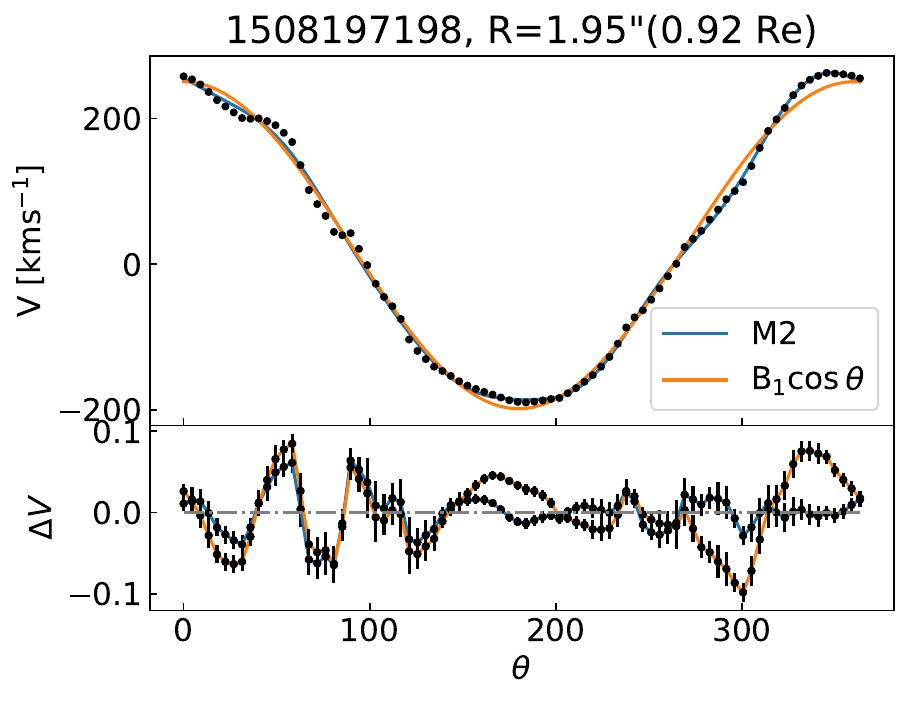}
    \caption{V$_{rot}$ as a function of azimuthal angle ($\theta$) around the ellipse for MAGPI1508197198. $\Delta$V are the residuals between the data and the model normalised to the maximum velocity of the galaxy. There are large (10s km s$^{-1}$ at $\theta$=50 and $\theta$=90) differences between the model and data that are not explained by the errors in the velocity. These differences are localised to specific regions of the galaxy, hence could be the result of local physical processes (like star formation) causing disturbances.}
    \label{Local_Dist}
\end{figure}

When finding the best fitting ellipse, \textsc{kinemetry} applies equal weighting to each spaxel along the ellipse, which essentially models a best-fitting `global' ellipse for that radius. Occasionally, there will be differences between the `global' model and the data that is not explained by the uncertainty in the velocity at that spaxel, implying these differences between the model and the data are located at a specific location within the galaxy; an example of this is shown in Fig. \ref{Local_Dist}. Including more higher order terms (i.e., the even terms) allows \textsc{kinemetry} to `catch' these local deviations from circular motion. Hence, these local deviations could be evidence of a local velocity disturbances (here we stress that this disturbance is different to the asymmetry, we are currently discussing). Since star formation is a local process, it becomes an obvious suspect for such a disturbance. Investigating the source of these disturbances will be done in subsequent work.

\subsubsection{Stellar Mass}
The result for low stellar mass, main sequence galaxies being more asymmetric could be evidence of star-formation feedback disturbing the kinematics. Stellar feedback is predicted to be a prominent source of feedback in low stellar mass galaxies \citep{2014MNRAS.445..581H,2018MNRAS.480..800H}; stellar winds and SNe would cause disturbances in the ionised gas kinematics (assuming the ionisation is from hot OB stars) that could induce asymmetries. The conditions that regulate star formation operate on a large span of scales (Mpc to pc), though it is accepted to be a largely local process (\citealt{2008AJ....136.2846B,2012ARA&A..50..531K}, but see \citealt{2020MNRAS.492...96B}). Depending on the spatial distribution of star formation, it could be responsible for localised kinematic disturbances in the disk of galaxies. 

\begin{figure*}
    \centering
    \includegraphics[scale=0.68]{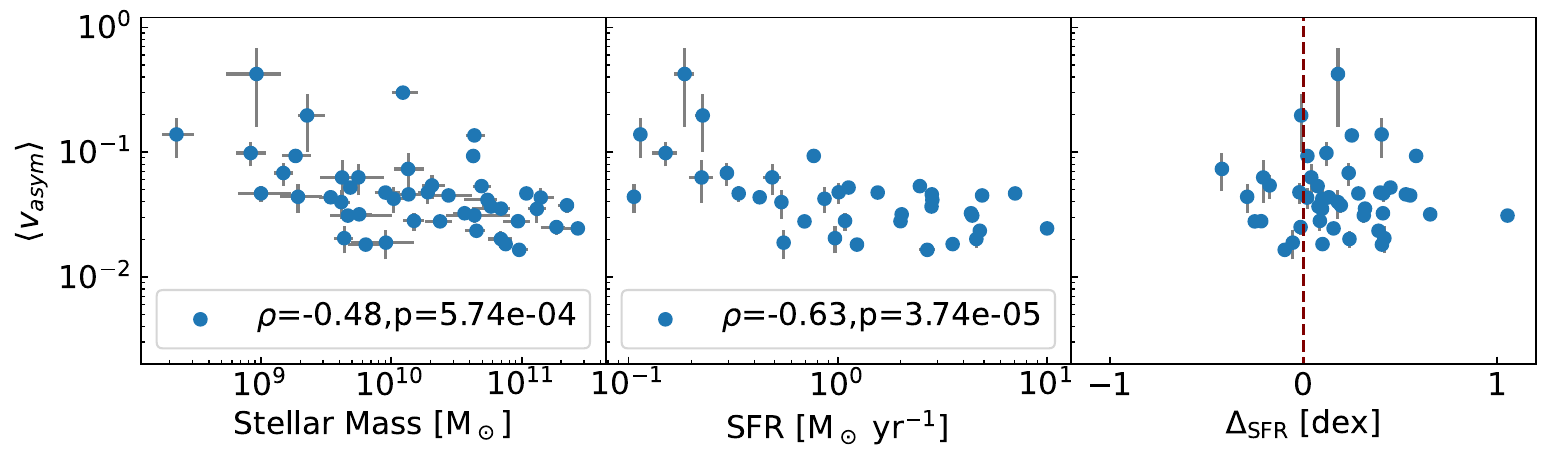}
    \caption{\textit{Left}: $\langle v_{\rm asym} \rangle$ vs. stellar mass, \textit{Centre}: $\langle v_{\rm asym} \rangle$ vs. star formation rate, \textit{Right}: $\langle v_{\rm asym} \rangle$ vs. $\Delta_{\rm SFR}$. The $\rho$ and p-values are calculated with respect to the $\langle v_{\rm asym}\rangle$. \HII+AGN galaxies are included in the left and right panel, but are excluded in the middle panel so as to investigate the relationship between purely star forming galaxies. Including \HII+AGN galaxies results in a considerably weaker correlation ($\rho=-0.36,p=0.00124$).} 
    \label{mass_asym}
\end{figure*}

A galaxy's stellar mass correlates with its rotational velocity \citep[i.e., The Tully-Fisher Relation;][]{1977A&A....54..661T}, hence we expect the asymmetry to decrease with increasing stellar mass. In agreement with \cite{2017MNRAS.465..123B} and \cite{2020ApJ...892L..20F}, we find that there is a relatively weak anti-correlation between asymmetry and stellar mass as shown in Fig. \ref{mass_asym}. We find that the strength of the anti-correlation between asymmetry and stellar mass is not affected by whether we use $v_{\rm asym}$ at $0.5R_e$ ($\rho = -0.41$, $p=0.00435$), $1.5R_e$ ($\rho = -0.54$, $p=0.00005814$) or $\langle v_{\rm asym} \rangle$ ($\rho = -0.48$, $p=0.000574$).

The absence of a strong anti-correlation with mass suggests that the connection between disturbed kinematics and a galaxy's stellar mass is not as straightforward as initially suggested in \cite{2002AJ....123.2358K} and \cite{2014ApJ...795L..37C}. A galaxies' SFR is strongly correlated with its stellar mass, and we find that $\langle v_{asym} \rangle$ has a stronger anti-correlation with SFR ($\rho=-0.63,p=0.0000374$) compared to stellar mass. A stronger anti-correlation could imply that the more non-circular motion the gas is exhibiting, the less gas is available for star-formation. Stellar/SNe feedback could be source for this non-circular motion/disturbed velocity, as we have discussed above.

Our sample consists of star-forming galaxies with S\'{e}rsic indices $\sim$ 1, which implies we are mostly looking at late-type galaxies. Late-type galaxies often host non-axisymmetric features such as bars \citep{2004ARA&A..42..603K,2013ApJ...779..162C}. Bars can cause gaseous inflows onto the nucleus of the galaxies, and these inflows would cause complex kinematics, particularly so for the inner regions. \cite{2022ApJ...939...40L} used a similar method to \textsc{kinemetry} to compute circular rotating models of barred galaxies that allowed for non-circular motions. They found evidence of velocity disturbances (characterised by higher-order Fourier coefficents) being caused by a non-axisymmetrical potential primarily in the the inner regions where the bar is dominant; and although there is very little gas in bars, they still had a noticeable effect on the ionised gas kinematics. It seems reasonable that a bar would be effective source for complex inner kinematics. Future work will be done to identify asymmetric galaxies with bars based on MAGPI's morphological classification (Foster et al., in prep.).

\section{CONCLUSIONS}
\label{sec::conc}
We present a study of the kinematic asymmetries present in the ionised gas of a 47 star-forming galaxies (0.27 $<z<$ 0.35) from the MAGPI survey to study how the kinematic asymmetry changes between the inner regions ($0.5R_e$) and outer regions ($1.5R_e$) of galaxies. We do this by fitting a series of ellipses to the line-of-sight velocity field using \textsc{kinemetry} and we quantify the asymmetry using similar practices in the literature. We investigate the different physical properties, in particular the environment, AGN, star formation activity, and stellar mass of our sample to find possible sources of the asymmetry in the velocity fields. 

Our findings are as follows:
\begin{itemize}
    \item After testing three different models, we find that the kinematics are sufficiently complicated to warrant the inclusion of even terms in the fitted Fourier Series. We find that roughly 56 \% of our sample displays asymmetric velocity fields ($\langle v_{\rm asym} \rangle > 0.04$); while 81 \% displays larger asymmetry at $1.5R_e$ compared to $0.5R_e$ and 46 \% are asymmetric at both radial extents. The large fraction of galaxies with more disturbed outskirts is consistent with kinematics being largely driven by the gravitational potential, which would be significantly shallower in the outskirts compared to the inner regions, and so the asymmetry could be larger at a further galactocentric radius.
    \item We find that most of our asymmetric galaxies belong in group environments ($N_{gals}<6$) where they would be affected by interactions from neighbouring galaxies. Focusing on the central galaxies, we find consistent results with interacting galaxies studied in simulations and previous studies using \textsc{kinemetry}, which found large asymmetries in interacting or merging galaxies. In line with other IFS surveys, we find that local environment, such as projected distance to nearest neighbour $d_1$, plays a role in whether a galaxy is asymmetric and that the most asymmetric galaxies tend to be satellites with small $d_1$.
    \item We find no evidence to suggest that AGN confirmed with a BPT diagram within 1.5$R_e$ preferentially drive inner asymmetries. The confirmed AGN hosts, as well as \HII+AGN galaxies, demonstrate significant asymmetry at both probed radii, which suggests that outflows (if they are present) or shocks disrupts the kinematics across the whole galactic disk. A larger sample of confirmed AGN host galaxies is needed to better understand the effect of AGN on the asymmetry. 
    \item For our sample, asymmetric galaxies with low stellar masses ($\log$ (M$_*$/M$_\odot$) $<$ 10) are preferentially found on the main sequence. We interpret this as possible evidence of stellar feedback driving the asymmetry in low stellar mass galaxies since stellar winds and SNe are primary sources of feedback in low stellar mass galaxies.
    \item We find a negative trend between $v_{\rm asym}$ and stellar mass, consistent with results from other IFS surveys. The negative trend persists, but does not increase in strength regardless whether $v_{\rm asym}$ is measured at $0.5R_e$, $1.5R_e$ or if we use a luminosity-weighted value. The negative trend between SFR and $v_{asym}$ is stronger than what was found with stellar mass. We suggest that the weak trend is possibly due to a non-uniform distribution of mass or non-axisymmetric feature present in the optical disk, and that stellar feedback is responsible for the stronger anti-correlation with SFR.
\end{itemize}

Kinematic asymmetries provide an excellent probe into physical processes that govern galaxy evolution, but the connection between the two is, evidently, complex. We can leverage the higher lookback time of MAGPI against IFS surveys in the local Universe to investigate whether there is any evolution in these kinematic asymmetries. In future work, we plan to apply a consistent analytical framework across different surveys and investigate asymmetries across a range of lookback times.  

\begin{acknowledgement}
We thank the referee for their constructive feedback on this manuscript draft. 
Based on observations collected at the European Organisation for Astronomical Research in the Southern Hemisphere under ESO program 1104.B-0536. We wish to thank the ESO staff, and in particular the staff at Paranal Observatory, for carrying out the MAGPI observations. 
MAGPI targets were selected from GAMA. GAMA is a joint European-Australasian project based around a spectroscopic campaign using the Anglo-Australian Telescope. GAMA is funded by the STFC (UK), the ARC (Australia), the AAO, and the participating institutions. GAMA photometry is based on observations made with ESO Telescopes at the La Silla Paranal Observatory under programme ID 179.A-2004, ID 177.A-3016. 
Part of this research was conducted by the Australian Research Council Centre of Excellence for All Sky Astrophysics in 3 Dimensions (ASTRO 3D), through project number CE170100013. 
CF is the recipient of an Australian Research Council Future Fellowship (project number FT210100168) funded by the Australian Government. CL, JTM and CF are the recipients of ARC Discovery Project DP210101945.
EW acknowledges support by the Australian Research Council Centre of Excellence for All Sky Astrophysics in 3 Dimensions (ASTRO 3D), through project number CE170100013.
LMV acknowledges support by the German Academic
Scholarship Foundation (Studienstiftung des deutschen Volkes) and the
COMPLEX project from the European Research Council (ERC) under the
European Union’s Horizon 2020 research and innovation program grant
agreement ERC-2019-AdG 882679.
YP acknowledge National Science Foundation of China (NSFC) Grant No. 12125301, 12192220, 12192222, and the science research grants from the China Manned Space Project with No. CMS-CSST-2021-A07.
PS is supported by Leiden University Oort Fellowship and the IAU Gruber Foundation Fellowship.
This work makes use of colour scales chosen from \citet{2020JOSS....5.2004V}. 
\end{acknowledgement}


\bibliography{bib}

\appendix

\section{Quantifying Kinematic Asymmetry}
\label{app1}
This appendix contains a description of how we deal with artefacts in the data, the \textsc{kinemetry} models used, comparison between the models and various tests we employed to ensure we chose the best model for our analysis.

\subsection*{Numerical workaround for gaps in the kinematic maps}
\begin{figure*}
    \centering
    \includegraphics[scale=0.7]{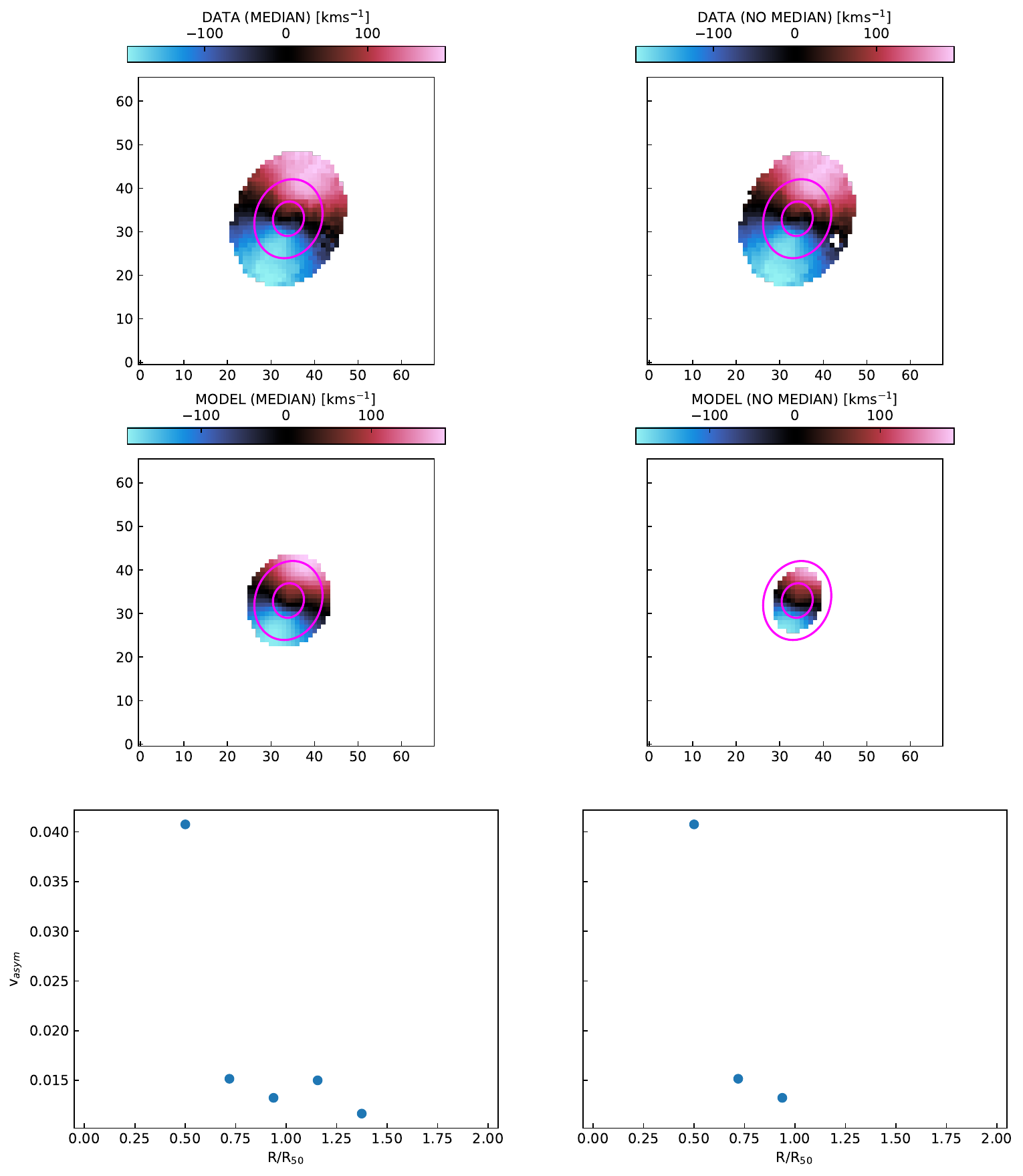}
    \caption{\textit{Top Left}: Velocity map for MAGPI1209197197 where we have masked the non-detected spaxels with median of the adjacent spaxels. \textit{Top Right}: Velocity map for MAGPI1209197197 where we have masked non-detected spaxels and not replaced them with the median of adjacent spaxels. \textit{Middle Left}: The fitted model from \textsc{kinemetry} from the velocity map in \textit{Top Left}. \textit{Middle Right}: The fitted model from \textsc{kinemetry} from the velocity map in \textit{Top Right}. \textit{Bottom Left}, $v_{asym}$ radial profiles from \textit{Middle Left}. \textit{Bottom Right}: The $v_{asym}$ radial profile from \textit{Middle Right}. Notice that when non-detected spaxels are not replaced, the fitted model does not have coverage out to $1.5R_e$, and we do have $v_{asym}$ values at $1.5R_e$, even though we have kinematic information at that radial extent.}
    \label{Non_detect_spaxels}
\end{figure*}

Our signal-to-noise cut sometimes rejects spaxels with otherwise valid neighbouring spaxels. These gaps cause the \textsc{kinemetry} algorithm to return trivial results for the corresponding radial ellipse, preventing a measure of the asymmetry at that radius and beyond even when data would otherwise allow. To mitigate this numerical artefact, we replace the velocity measurement of rejected spaxels with the median value from their eight adjacent spaxels to ensure we are not removing galaxies where we have kinematics out to our radii of interest (See Fig. \ref{Non_detect_spaxels}. Ellipses where we have thus replaced a significant fraction (more than 30\%) of the spaxels along the ellipse with medians are exclude from further analysis. Hence, we can be sure that this numerical workaround does not leave a galaxy with poor \textsc{kinemetry} fits in our sample, and it does not remove a galaxy where we do indeed have a good fit.

\begin{figure*}
  \centering
  \subfigure{\includegraphics[scale=0.55]{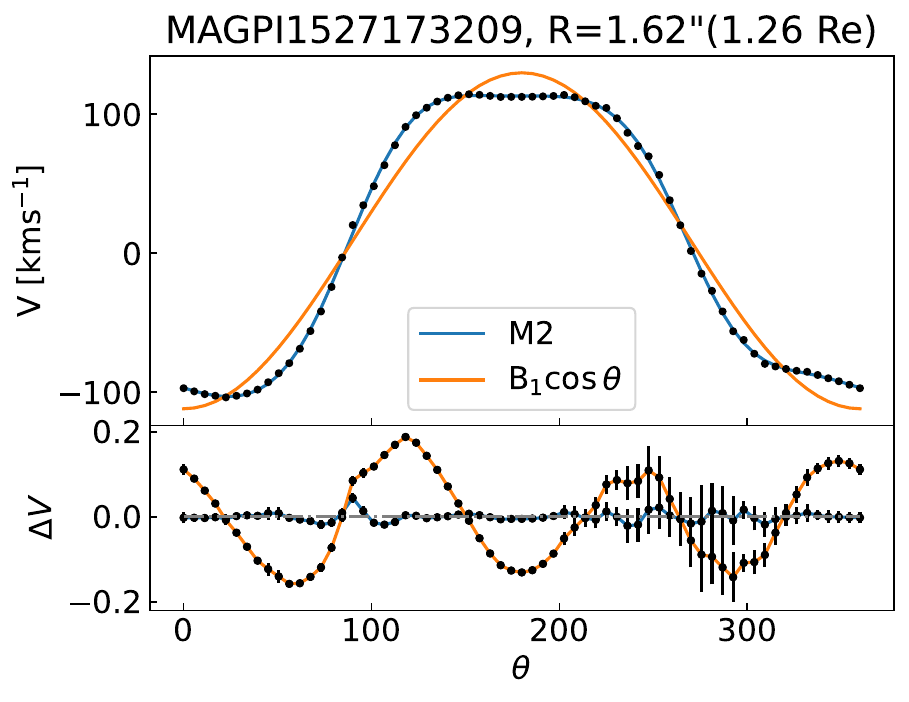}}
  \subfigure{\includegraphics[scale=0.55]{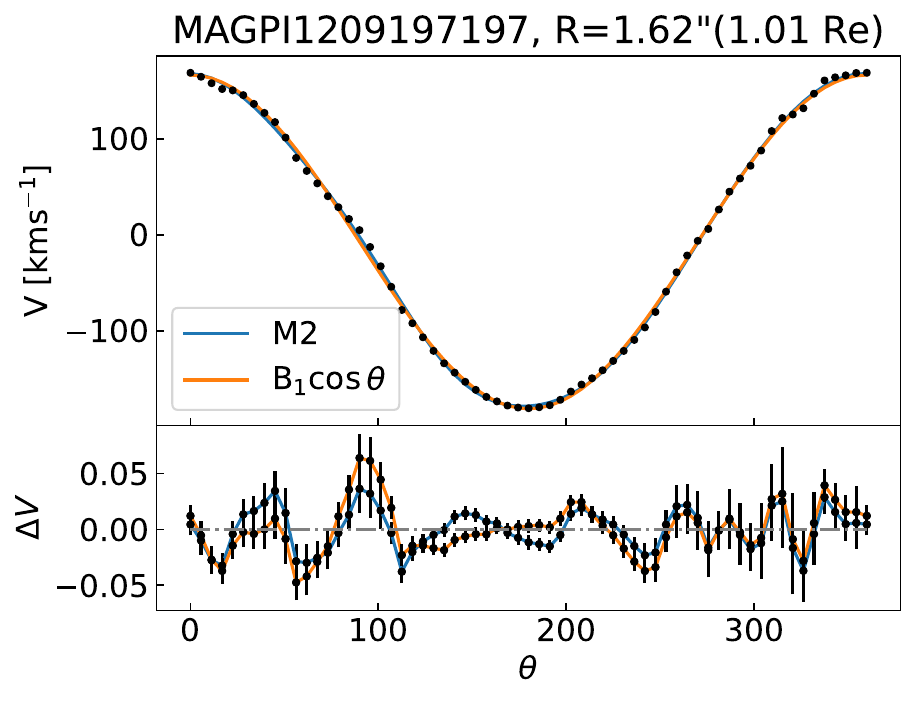}}
  \caption{Rotational velocity (V$_{rot}$, top) and residuals (bottom) as a function of azimuth around the ellipse with semi-major axis of 1.62$''$ for MAGPI1527173209 (left) and MAGPI1209197197 (right). The orange line shows the expected values from a simple circular rotating model (ie. $B_1\cos(\theta)$). The blue line is the fitted M2 model, which includes the higher order Fourier coefficients. Both velocity profiles can be adequately explained by M2. MAGPI1209197197 has a very symmetrical profile, whereas MAGPI1527173209 has asymmetric features that are not explained by the uncertainties (shown as errorbars in the bottom panel), examples of which can be seen around $\theta$ = 100$^\circ$ and 300$^\circ$ in the bottom left panel. Also note that these differences are an order of magnitude larger than those in the symmetric galaxy}
  \label{ellipse_plots}
\end{figure*}

\subsection*{Comparison between models}
\begin{figure*}
  \centering
  \includegraphics[scale=0.47]{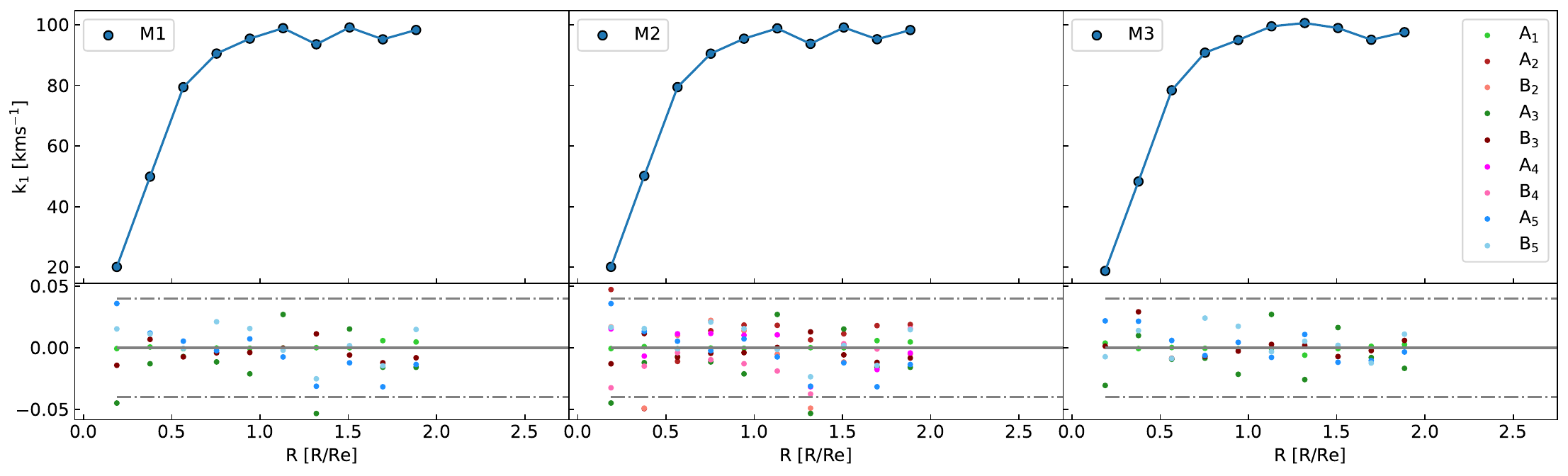}
  \caption{Radial profiles of \textsc{kinemetry} coefficient from each model for galaxy MAGPI1205197197, a particularly well-resolved galaxy where we could fit 10 ellipses. The higher order coefficients, which have been normalised to $B_1$, are plotted below. The left panel is M1, the middle panel is M2 and the right panel is M3. The middle panel includes terms that are not listed in the legend, but match the colors in the legends on either side. The grey dashed lines are the 0 and $\pm$0.04 \% asymmetric limits commonly adopted in the literature. The solid grey line represent zero. Having a variable centre slightly decreases the $k_n$ values, whereas including even terms does not. The even terms are also a similar order of magnitude as the odd terms, and do not decrease the value of the odd terms.}
  \label{nterms_FS}
\end{figure*}

\cite{2006MNRAS.366..787K} warned that adopting incorrect ellipse geometry (i.e. centre coordinates, PA and $q$) can result in additional power being distributed to other coefficients. In particular, an incorrect centre will cause $A_0, A_1, A_2, B_2, A_3$ and $A_4$ to have amplified values. This causes large, and in cases azimuthally periodic, residuals to appear between the data and the model that need to be considered in order to maintain a good fit to the data. This was noticed when running M1 on the MAGPI velocity maps, and motivated us to explore M2 and M3 as potential options.

\subsubsection*{Model 1: odd terms only, fixed centre}
\begin{figure}
    \centering
    \includegraphics[scale=0.55]{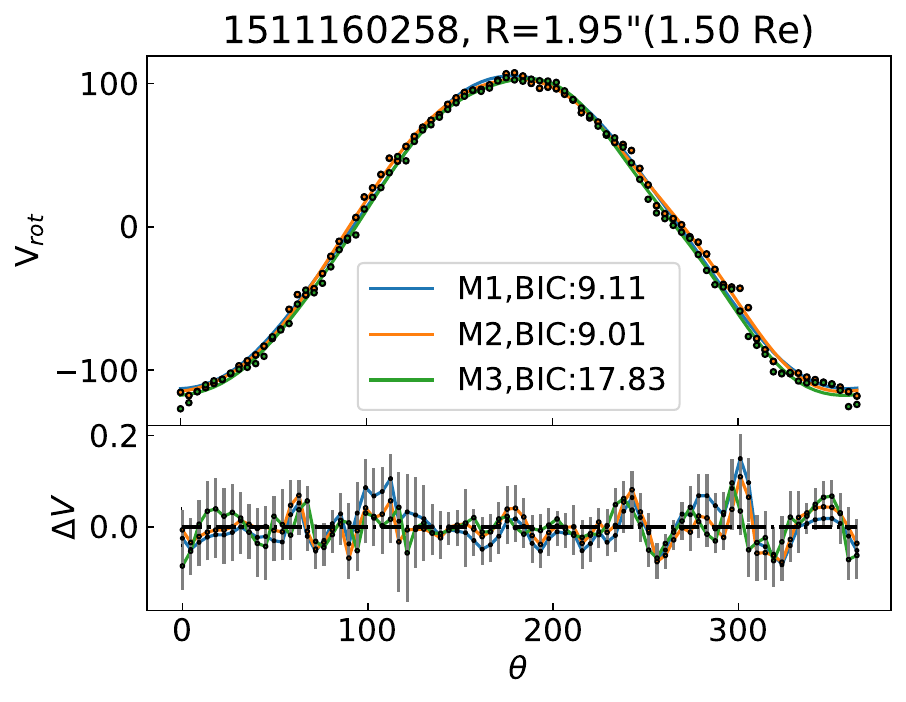}
    \caption{The fitted \textsc{kinemetry} ellipse from M1 (blue), M2 (orange) and M3 (green). The bottom panel shows the residuals from the fit, normalised to the maximum V$_{\rm rot}$ at that ellipse. The errorbars shown are the errors on the velocities at those spaxels. The BIC values for the fitted ellipse are shown. Both M2 and M3 provide better fits to the data, suggesting that either an incorrect kinematic centre is being used, or even terms are required to ensure a good fit to the data.}
    \label{M1_M2_M3_ellipse}
\end{figure}

For all galaxies in our sample, we find a residual sinusoidal pattern with azimuth between our data and M1. An example of this sinusoidal residual pattern is shown in the left panel of Fig. \ref{ellipse_plots}, where a significant and systematic difference is seen between the data and a simple circular rotating model as a function of azimuth. These periodic residuals imply that the model is unable to accommodate additional harmonic component(s) in the fitting. \textit{Due to these systematic residuals in M1, we do not use M1 in subsequent analyses.}


\begin{figure}
    \centering
    \includegraphics[scale=0.55]{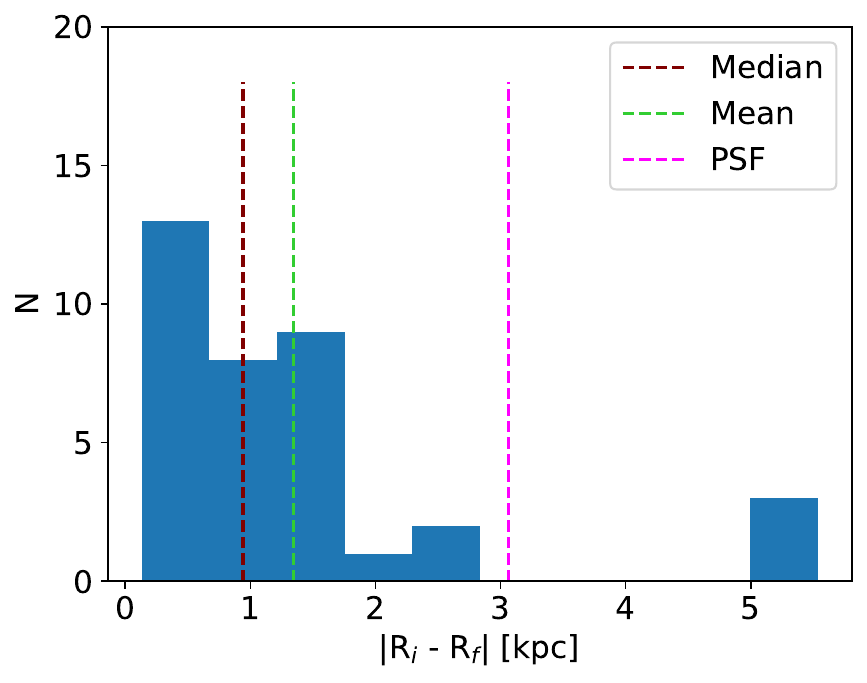}
    \caption{Histogram of the distance between the original centre (R$_i$) coordinates and the final fitted centre (R$_f$) for M3. The median and mean are shown as maroon and green lines. The HWHM of the estimated PSF is shown in magenta. In most cases, the distance between initial and final coordinates is less the PSF. This implies that the the movement of the centre may not be physical, or at least we cannot be sure if its real at our current resolution limit.}
    \label{r_dist}
\end{figure}

\subsubsection*{Model 3: odd terms only, variable centre}
To determine which model is to be preferred between M2 and M3, we first determine if the power to the even terms in M2 is similar to that being distributed to the odd terms in M3. We do this by plotting radial profiles of each coefficient. Fig. \ref{nterms_FS} shows radial value of each coefficients ($A_2,B_2, A_3$ etc...) for a typical galaxy in our sample. For most of the galaxies that are rotating as expected (ie. low asymmetry), the power being distributed to the even terms is of similar magnitude to the odd terms at all radii. The galaxies that display large even coefficients in M2 also have large odd coefficients, and have very disturbed velocity maps (i.e., they would be considered asymmetric even if using odd terms). The comparable magnitude of the even terms present in galaxies with regular velocity maps and disturbed velocity maps suggests that the even terms are, in fact, needed to adequately model not only the particularly disturbed velocity maps for our sample, but those galaxies which are rotating regularly as well. Including even terms, however, does not decrease the power being distributed to odd terms that could be affected by an incorrect ellipse geometry, suggesting that while it does improve the quality of the fit, it does not eliminate the amplification of $A_1, A_3, B_3$.

\begin{figure}
    \centering
    \includegraphics[scale=0.55]{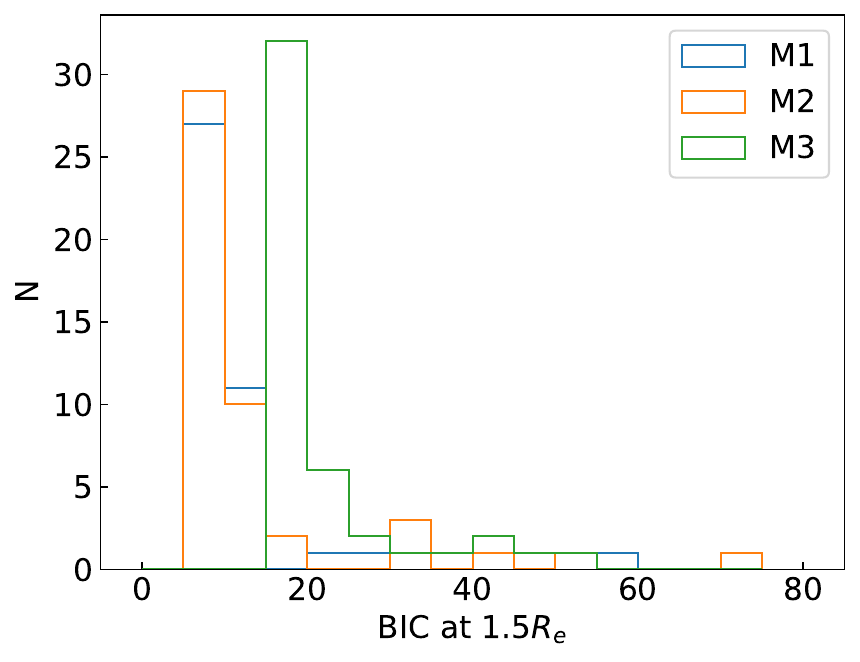}
    \caption{Histogram of BIC values for the fitted \textsc{kinemetry} ellipse $1.5R_e$ for galaxies in our sample. The BIC values for M3 are consistently higher for every galaxy than both M1 and M2. Occasionally, M1 will have a larger BIC values than M2, indicating that M2 is better model, given our data.}
    \label{BIC_plot}
\end{figure}

\begin{figure*}
    \centering
    \includegraphics[scale=0.7]{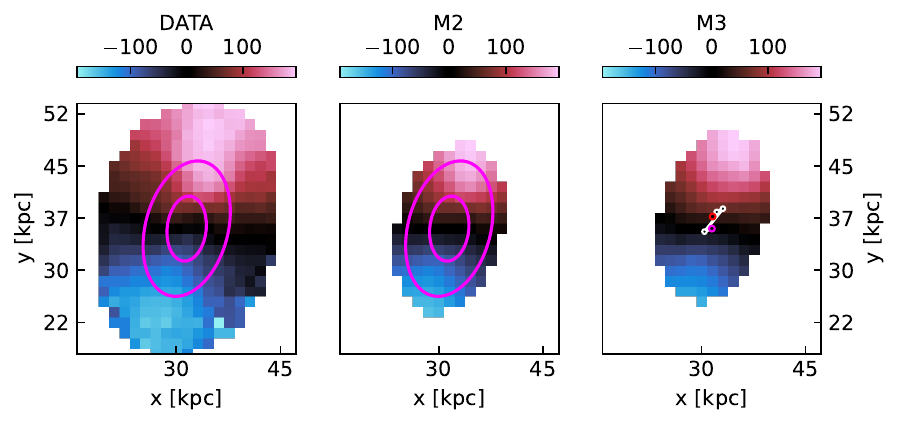}
    \caption{The velocity map, and fitted \textsc{kinemetry} models, for the asymmetric galaxy MAGPI1503208231. The left panel is the data, the middle panel is fitted model from M2, and the right most panel is the fitted model from M3. The black and white dots are the coordinates of the best-fitting centre. The red outline indicated the final ellipse's centre coordinates. The magenta outline indicates the original coordinates. The size of the estimated PSF is shown in the bottom left corner. In most cases, the end coordinate returns sufficiently close to the original centre. It is important to notice that travelling kinematic centre hardly ever reaches distances larger than a PSF, suggesting that change is not physical meaningful. Moreover, the variable centre goes on a random walk around the original centre.}
    \label{M2_M3_washout}
\end{figure*}

\textsc{kinemetry}, like most tilted-ring fitting algorithms, works with a degenerate parameter space, especially when allowing PA$_{\rm kin}$, $q_{\rm kin}$ \textit{and} the centre coordinates to vary \citep{2010A&A...516A..11L,2015MNRAS.451.3021D}. Fig. \ref{nterms_FS} shows that allowing the centre's coordinates to vary does decrease the power distributed to the odd terms (i.e., $A_1, A_3, B_3$) and also improves the quality of the fit to the same level as the inclusion of even terms (i.e., residuals for M2 and M3 are similar). However, a variable kinematic centre could prove problematic if \textsc{kinemetry} begins to `overfit' to the data. To gauge this, we calculate a Bayesian Information Criterion \citep[BIC;][]{2007MNRAS.377L..74L} for each ellipse. The BIC can be computed by
\begin{equation}
    BIC = \chi^2 + k\ln N
\end{equation}
where $\chi^2$ is the traditional chi-squared-goodness-of-fit value (or maximum liklihood with Gaussian errors, in a Bayesian context), $k$ is the number of free parameters and $N$ is the number of data used in the fitting. The advantage in using a BIC over a reduced $\chi^2$ is that unnecessary parameters are penalised. The middle and right-most panel in Fig. \ref{M2_M3_washout} shows the 2D model velocity maps from M2 and M3. What \textsc{kinemetry} fits is a series of ellipses, and creates a 2D model by interpolating between ellipses. To see which model (M1, M2 or M3) is the best representation, given our data, we calculate the BIC value at $1.5R_e$, the same ellipse we use to exclude galaxies in Fig. \ref{v_asym_SNR}. Fig. \ref{BIC_plot} shows the distribution BIC values for galaxies in our sample. M3 consistently has larger BIC values than both M1 and M2, while occasionally M1 will have larger BIC values. This suggests that M3 overfits our data and is therefore not our preferred model.

We also investigate where the variable kinematic centre travels on the map by calculating how far it during the fitting routine. When we look at the distance the kinematic centre travels for M3, we find that the centre of most galaxies only moves roughly one spaxel (0.2 kpc) from the original coordinates with mean and median of 1.1 (1.42 kpc) and 0.96 (1.21 kpc) spaxels, respectively. At the maximum redshift of our sample ($z=0.35$), the physical scale of the spatial PSF corresponds to $\sim$ 3 kpc (1.65 spaxels). Of the 47 galaxies fitted with M3, only 4 have a greater maximum displacement than 3 kpc. Furthermore, in most cases, the kinematic centres return to the original coordinates as the number of spaxels fitted increases with radius (shown in Fig. \ref{M2_M3_washout}), suggesting that the centre displacement does not correspond to a real physical movement of the kinematic centre. Given our findings that allowing for varying kinematic centre seems to have a marginal effect on the position of the centre compared to the PSF, \textit{we do not continue to use M3.}

\subsubsection*{Model 2: inclusion of even terms}
Including even terms could cause the asymmetry to be inflated by virtue of including additional components in the sum in Eqn. \ref{SINS}. To investigate this, we split $v_{\rm asym}$ in its even and odd components using the equations below,

\begin{multicols}{2}
  \begin{equation}
    v_{\rm asym,even}=\frac{k_2+k_4}{2k_1},
  \end{equation}\break
  \begin{equation}
    v_{\rm asym,odd}=\frac{k_3+k_5}{2k_1}.
  \end{equation}
\end{multicols}

\begin{figure}
    \centering
    \includegraphics[scale=0.55]{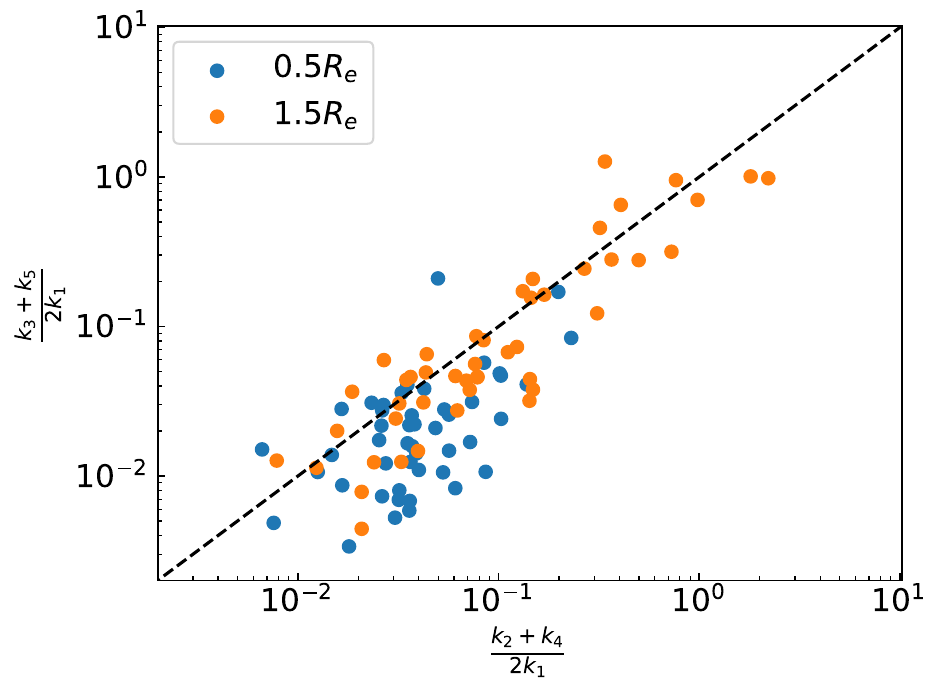}
    \caption{Comparison of the power to the even ($\frac{k_2+k_4}{2k_1}$) vs. odd ($\frac{k_3+k_5}{2k_1}$) moments for M2. The black dot dashed line represents the one-to-one. For low values and within $0.5R_e$ (orange symbols), even and odd terms display similar contributions to the asymmetry; however at $1.5R_e$ (blue symbols), the odd terms dominate the asymmetry measure.}
    \label{even_cont}
\end{figure}

The odd and even contributions are shown in Fig. \ref{even_cont}. We find that for small values of asymmetry ($v_{\rm asym,odd}$, $v_{\rm asym,even}<0.04$ ie. low asymmetry) the even terms are comparable to the odd terms. The even terms appear particularly important at $0.5R_e$ where they scatter above and below the 1:1 line. At $1.5R_e$, the odd terms dominate $v_{\rm asym}$, but still with a non-negligible component in the even terms (ie. are still on the 1:1 line). We conclude that even terms are necessary to adequately fit our galaxy sample and must be included.

After much deliberation and testing, we completed our analysis using M2.


\end{document}